\documentclass[preprint,flushrt]{aastex}

\usepackage{amsmath}
\usepackage{amssymb}
\usepackage{amsfonts}

\newcommand{\sinc}{\textrm{sinc}}
\newcommand{\bs}[1]{\boldsymbol{#1}}
\newcommand{\bsrm}[1]{\mathbf{#1}}

\begin{document}

\title{Characterizing the Galactic Gravitational Wave\\ Background
  with LISA}

\author{Seth E. Timpano\altaffilmark{1,2}, Louis
  J. Rubbo\altaffilmark{2}, Neil J. Cornish}

\affil{Department of Physics, Montana State University, Bozeman, MT
  59717}

\altaffiltext{1}{Current address: Department of Physics, Pennsylvania
  State University, University Park, PA 16802}

\altaffiltext{2}{Current address: Center for Gravitational Wave
  Physics, Pennsylvania State University, University Park, PA 16802}

%#### MAIN DOCUMENT ###############################

%==== Abstract ====================================

\begin{abstract}
We present a Monte Carlo simulation for the response of the Laser
Interferometer Space Antenna (LISA) to the galactic gravitational wave
background.  The simulated data streams are used to estimate the
number and type of binary systems that will be individually resolved
in a 1-year power spectrum.
We find that the background is highly non-Gaussian due to the presence
of individual bright sources, but once these sources are identified
and removed, the remaining signal is Gaussian.  We also present a new
estimate of the confusion noise caused by unresolved sources that
improves on earlier estimates.
\end{abstract}

%==== Keywords ====================================


%==== Text ========================================

\section{INTRODUCTION} \label{sec:intro}

Binary star systems are excellent sources of gravitational waves, and
with roughly two-thirds of the $\sim$$10^{11}$ stars in our galaxy in
binary systems, there will be no shortage of targets for the proposed
Laser Interferometer Space Antenna (LISA) \citep{LPPAR}.  Binaries
with periods less than a day may potentially dominate the response of
the LISA observatory.  Indeed, it is likely that the main source of
noise for LISA over a portion of its band will be unresolved
gravitational wave signals from galactic and extra-galactic binary
star systems. Several studies \citep{EIS87, HBW90, BH97, PP98, NYZ01}
have sought to model these populations and estimate their contribution
to the gravitational wave power spectrum. In some instances these
estimates have been combined with data analysis considerations to make
predictions of the confusion noise caused by unresolved sources.
Although a consensus has not formed on the expected background level,
it is generally accepted that a galactic gravitational wave background
does exist inside the LISA band.  The difficultly in developing strong
limits on the background level originate from the expectation that the
background will be dictated by compact binaries, that is, binary
systems that contain white dwarfs, neutron stars, and stellar mass
black holes.  Since the electromagnetic luminosity of compact binaries
is low, not enough sources have been observed yet to build reliable
models for the populations.

If current estimates of compact binary populations reasonably
represent the true nature of the galaxy, then the superposition of
gravitational wave signals from these populations will form a
confusion limited background in the LISA band.  That is, there will be
enough sources that the received signals will interfere with each
other to the point where individual binaries cannot be
resolved~\citep{CC04}.  When this occurs the background becomes a
source of noise in the detector.  However, around ten thousand systems
will be resolvable due to either their isolation in frequency space
(for sources with frequencies above $\sim$3~mHz) or their relative
brightness (for sources below $\sim$3~mHz).

Here we present a Monte Carlo model for the galactic gravitational
wave background.  Our goal is to better understand the role played by
the rare, bright sources that dominate the observed signal, and to
provide a more realistic level of the confusion background due to
unresolved compact binaries.  Our investigation of the galactic
gravitational wave background is done in two phases.  The first phase
is to build a Monte Carlo simulation of the background by modeling
each binary and processing the corresponding gravitational wave signal
through a model of LISA.  To reasonably represent the individual types
of binaries we follow the population models presented in
\citet{HBW90} and \citet{NYZ04}, which from here on will be referred
to as HBW and NYZ respectively. Though less up to date, the HBW model
has the advantage of being expressed in terms of explicit distribution
functions, which allows us to generate multiple realizations. The more
modern NYZ model employs a population synthesis code that has not been
made public, so we were not able to generate our own realizations. Gils
Nelemans was kind enough to send us a realization of the NYZ model
to work with. The second phase of our study is to statistically characterize
features of the galactic background as they are observed by LISA.

As part of the modeling phase, the signal from each source is run
through a realistic model of the LISA instrument response to produce
simulated interferometry data.  In doing so, it was necessary to develop a new,
fast algorithm for computing the detector response in order to process
the $\sim$$10^8$ sources modeled in each realization. The algorithm is
based on the frequency domain approach developed by \citet{CL03b}, and
extended to include the detector transfer functions, arbitrary
observation times, and frequency evolution of the sources. For the
current study we set the observation time at $T_{\rm obs} = 1\; {\rm year}$.
Examples of our simulated LISA data streams can be found at the
\textit{Mock LISA Data Archive}\footnote{\texttt{http://astrogravs.gsfc.nasa.gov/docs/mlda/}}.

Using the simulated detector data we investigate issues that are of
importance to the development of future data analysis algorithms.
Among the quantities we investigate are tests of Gaussianity in the
distribution of Fourier coefficients before and after bright sources
are removed, the number and type of bright sources, and the density
(in frequency space) of bright sources.  Our interest in bright
systems stems from the idea that they will be identifiable in the data
streams, and thus removable.  They will also be instrumental in using
the real gravitational wave data to study galactic populations and
galaxy evolution.

To model the removal of bright systems we use an iterative procedure
using a running median of co-added instrument noise and galactic
signals as the effective noise level. Sources were considered bright
if they had a signal noise ratio (SNR) greater than some threshold 
with respect to the effective noise level. We considered both
optimistic (${\rm SNR}=5$) and conservative (${\rm SNR}=10$) thresholds.
As the bright sources are regressed from the data, the effective noise
level drops, allowing more sources to be resolved. After several iterations
we are left with a residual signal that is our estimate of the galactic
confusion noise. Previous estimates of the confusion noise were derived
by setting a maximum source density, with the reasoning that it would
be impossible to resolve individual sources when the number of sources
per $1/T_{\rm obs}$ frequency bin exceeded some threshold.
Here we took a different
approach that is based primarily on SNR thresholds, but we also studied
the effect of a source density cut-off. Rather than working with the
total source density we based our cut-off on the density of resolved
sources. In other words, we considered the possibility that there will
be a maximum number of sources that can be resolved per frequency bin.
We studied the effect of a resolved
source density cut-off by performing the iterative removals with and
without a cut-off on the number of sources that could be
resolved per frequency bin (we set a limit of one source per four
bins). For some models the source density cut-off had a significant
impact, but for other models the cut-off made very little difference.
Our estimate of the confusion level for the HBW model differs
from that of \citet{BH97} despite the fact that we use the same galactic
model. Our estimate of the confusion level for the NYZ model agrees
fairly well with the \citet{BC04} estimate. In both cases, our estimate
is lower at low frequencies (below 1 or 2 mHz respectively), and higher at high
frequencies.  The differences are due to our differing
approaches to modeling the signal identification and regression. We
feel that our approach yields more realistic estimates. All our examples
are for one year of observations, so the frequency bins have width
$\Delta f = 3.17 \times 10^{-8}$ Hz. The level of the confusion background
will drop for longer observation times as the sidebands get better resolved
and the SNR increases. The reduction in the confussion noise over time
means that fits to transient sources that occur in the first year of operation,
such as a supermassive black hole merger, will continue to get better with
time even though the source disappeared years ago!

Recently \citet{BDL04} and \citet{ETKN05} simulated LISA time series
for a population of galactic white dwarf binaries.  While comparable
to our approach, neither simulation was used to study source
identification and subtraction. The statistical analysis of the
background given in \citet{ETKN05} focuses on the cyclostationary
nature of the signal, whereas our statistical analysis focuses on
simulating data analysis in order to better understand the galactic
gravitational wave background.

Since the study of the galactic gravitational wave background
naturally divides itself into two sections, modeling and
characterization, the outline of the paper follows suite.
Sections~\ref{sec:model} and \ref{sec:detector_background} are devoted
to a description of the Monte Carlo simulation of the galactic close
binary populations.  It is here that we describe how the individual
sources are modeled and convolved with a LISA response model.  The
next three sections calculate a number of statistical properties
associated with the galactic background.  Section~\ref{sec:gauss}
demonstrates that the galactic background is non-Gaussian in nature.
In Section~\ref{sec:clb} we present our estimate of the confusion
limited background and compare it to prior estimates.
Section~\ref{sec:bright_sources} describes the characteristics of the
systems that are labeled as bright.  The paper concludes in
Section~\ref{sec:conclusion} with a discussion of the various
assumptions used in the simulation and how changes in these
assumptions may alter our results.

\section{GALACTIC MODEL} \label{sec:model}

The first step in building a Monte Carlo realization of the galaxy is
to model an individual binary system.  In general, a gravitational
wave traveling in the $\hat{k}$ direction can be decomposed into two
polarizations states,
\begin{equation}
  \bsrm{h}(c t - \hat{k}\cdot\bs{x}) = h_{+}(c t - \hat{k}\cdot\bs{x})
  \bs{\epsilon}^{+} + h_{\times}(c t - \hat{k}\cdot\bs{x})
  \bs{\epsilon}^{\times} \,,
\end{equation}
where $\bs{\epsilon}^{+,\times}$ are basis tensors used to describe
the radiation's orientation.  The scalar coefficients are referred to
as the gravitational waveforms.  For a circular binary with
instantaneous angular orbital frequency $\Omega$, and component mass
$M_{1}$ and $M_{2}$ the waveforms measured at the barycenter of the
Solar System are
\begin{mathletters} \label{eq:waveforms_background}
\begin{eqnarray}
  h_+(t) &=& A_+ \cos(2\psi) \cos(2\Omega t + \varphi_{o}) + A_\times
  \sin(2\psi) \sin(2\Omega t + \varphi_{o}) \\
  h_\times(t) &=& -A_+ \sin(2\psi) \cos(2\Omega t + \varphi_{o}) + A_\times
  \cos(2\psi) \sin(2\Omega t + \varphi_{o}) \,,
\end{eqnarray}
\end{mathletters}
where the polarization amplitudes are given by
\begin{mathletters} \label{eq:pol_amps}
\begin{eqnarray}
  A_{+} &=& \frac{2 G^2 M_1 M_2}{ c^4 r} \left( \frac{\Omega^2}{G(M_1
    + M_2)} \right)^{1/3} \big(1 + \cos^2(\iota) \big) \\
  A_{\times} &=& -\frac{4 G^2 M_1 M_2}{c^4 r} \left(
    \frac{\Omega^2}{G(M_1 + M_2)} \right)^{1/3} \cos(\iota) \,.
\end{eqnarray}
\end{mathletters}
The angles $\psi$ and $\iota$ describe the orientation of the binary
as viewed by an observer in the barycenter frame, while $\varphi_{o}$
is the initial phase.

Gravitational waves carry away energy and angular momentum from the
emitting system.  Consequently a binary will slowly inspiral over
time.  For stellar mass galactic sources in the LISA band the period
evolution can be adequately described by
\begin{equation} \label{eq:period_evolution}
  P_{orb}(t) = \left(P_{o}^{8/3} - \frac{256}{5 c^{5}}(2\pi)^{8/3} (G
  \mathcal{M})^{5/3} t \right)^{3/8} \,,
\end{equation}
where $\mathcal{M} \equiv (M_{1} M_{2})^{3/5} (M_{1}+M_{2})^{-1/5}$ is
the so-called chirp mass and $P_{o}$ is the initial orbital period.
Equation~\eqref{eq:period_evolution} does make the assumption, which
is used throughout this paper, that no other processes (e.g. mass
transfer) are evolved in the binary evolution besides gravitational
wave emission.

Equations~\eqref{eq:waveforms_background}, \eqref{eq:pol_amps}, and
\eqref{eq:period_evolution} indicate that a circular binary is
uniquely determined by a set of nine parameters: the component masses
($M_{1}, M_{2}$), initial orbital period ($P_{o}$), binary orientation
$(\psi, \iota$), initial phase ($\varphi_{o}$), and the distance to
the source ($r$).  Additionally, two angular variables ($\theta,
\phi$) are used to locate the source on the celestial sphere.  To
model an individual binary requires an accurate representation of
these nine parameters.

The list of source parameters are separable into those that are
extrinsic and intrinsic to the system.  The extrinsic variables $\{r,
\theta, \phi, \psi, \iota, \varphi_{o}\}$ do not influence the
evolution of the binary.  Instead they depend on the time of
observation and on the location of the observer with respect to the
binary.  The remaining variables $\{M_{1}, M_{2}, P_{o}\}$ directly
effect the binary evolution through the emission of gravitational
waves via equation~\eqref{eq:period_evolution}.

\subsection{Extrinsic Parameters}

For the set of extrinsic variables there is a further separation into
those that locate the source $\{r, \theta, \phi\}$ and those that
describe the time of observation and orientation as viewed by a
particular observer $\{\psi, \iota, \varphi_{o}\}$.  To derive a
unique location for each source we use a cylindrically symmetric disk
model of the galaxy with an exponential falloff in both the radial and
vertical directions,
\begin{equation} \label{eq:galactic_model}
  \rho = \rho_{o} e^{-r/r_{o}} e^{-|z|/z_{o}} \,.
\end{equation}
Here $\rho_{o}$ is the space density at the galactic center, $r_{o}$
is the radial scale length, and $z_{o}$ is the vertical scale height
of the galactic disk.  The values of $r_{o}$ and $z_{o}$ vary with the
different types of binaries (i.e. cataclysmic variables, white dwarf
binaries, etc.), but all types are assumed to obey the above model.
The binary positions are simply described in
galactocentric-cylindrical coordinates.  The natural coordinate system
for the LISA mission is heliocentric-ecliptic coordinates.  Therefore,
once the positions for the binaries are selected using the galactic
position distributions, they are translated to the LISA coordinate
system through a series of standard coordinate transformations.

The observed orientation of a binary system is set by the principal
polarization angle $\psi$ and the inclination angle $\iota$.  The
inclination angle is defined as the angle between the line of sight to
the binary $\hat{n}$ and the angular momentum vector of the binary
$\vec{L}$.  The inner product of $\hat{n}$ and the angular momentum
directions $\hat{n}\cdot\hat{L}$ is taken to be uniformly distributed
between $-1$ and $1$.  The principle polarization angle describes the
orientation of the semi-major axis of the projected binary orbit on
the celestial sphere and is uniformly distributed between 0 and $\pi$.
The distribution for $\varphi_{o}$, which describes the positions of
the binary components at time $t=0$, is uniformly distributed between
0 and $2\pi$.

\subsection{Intrinsic Parameters}

The distributions for each intrinsic parameter $\{M_{1}, M_{2},
P_{o}\}$ depends on the binary type under consideration.  To model
each of these parameters we used the distributions given in HBW.  For
this reason our galactic backgrounds include W~UMa ($3 \times
10^{7}$), cataclysmic variables ($1.8 \times 10^{6}$), neutron star -
neutron star ($10^{6}$), black hole - neutron star ($5.5 \times
10^{5}$), and close white dwarf ($3 \times 10^{6}$) binaries.  The
quantities in the parentheses indicates how many systems of that type
are included in the simulation.  Note that for most of our analyses we
use the 10\% reduced population of close white dwarf binaries as
described in HBW as this allows us to to compare directly with prior
results.

We have elected not to include the unevolved binaries.  The reason for
this is that they are predominately very low frequency sources
($\lesssim 10^{-5}$~Hz).  As a result, their signals will be buried in
the instrumental noise and, therefore, will not contribute to the
observed galactic background.

\subsection{Barycentric Background}

Many of the prior studies of the galactic background approached the
problem by estimating the net gravitational wave luminosity as a
function of frequency in the Solar System barycenter.  From the
luminosity they then derive a gravitational wave strain amplitude
using \citep{DB79}
\begin{equation}
  h = \left( \frac{16 \pi G}{c^{3} \omega_{gw}^{2}} \frac{L_{gw}}{4\pi
    r^{2}} \right)^{1/2} \,,
\end{equation}
where $\omega_{gw}$ and $L_{gw}$ are the gravitational wave angular
frequency and luminosity respectively.

In order to make comparisons between our results and prior studies, we
must relate the above expression for the strain amplitude to
quantities calculated in our Monte Carlo simulation.  To this end, we
first note the relationship between gravitational wave luminosity and
flux \citep{PT72},
\begin{equation}
  \frac{L_{gw}}{4 \pi r^{2}} = F_{gw} = \frac{c^{3}}{16\pi G}
  \big\langle \dot{h}_{+}^{2} + \dot{h}_\times^{2} \big\rangle \,,
\end{equation}
where the angle brackets denote an average over several gravitational
wave periods.  Using this relationship the strain amplitude is
rewritten as
\begin{equation}
  h = \left( \frac{1}{\omega_{gw}^{2}}  \big\langle \dot{h}_+^{2} +
  \dot{h}_\times^{2} \big\rangle \right)^{1/2} \;.
\end{equation}
From the waveforms given in equation~\eqref{eq:waveforms_background}
it follows that
\begin{equation} \label{eq:strain_background}
  h = \left( \frac{1}{2} (A_+^2 + A_\times^2 ) \right)^{1/2} \,.
\end{equation}
The polarization amplitudes, $A_+$ and $A_\times$, are functions of
the binary masses, distance to the source, orbital period, and
inclination angle (see eq.~[\ref{eq:pol_amps}]).

Equation~\eqref{eq:strain_background} gives the strain amplitude for a
single source.  To mimic a power spectrum in the Solar System
barycenter frame we first bin the sources according to their
frequencies.  The bin widths are $\Delta f = 1/T$, where $T$ is the
total observation time (for our simulations $T$ is set to one year).
Once the sources are sorted the net strain amplitude per frequency bin
is calculated from
\begin{equation} \label{eq:h_net}
  h_{net} = \left( \frac{1}{2} \sum_{i=1}^{N_{b}} \left( A_{+}^{2} +
  A_{\times}^{2} \right)_{i} \right)^{1/2} \,,
\end{equation}
where $N_{b}$ is the number of sources in the bin.  Note that
equation~\eqref{eq:h_net} accounts for constructive and destructive
interference in the same way as in standard data analysis
uncorrelated, random errors add quadratically with the square root
taken after all errors have been included.  Similarly for the
background, the net strain amplitude is the square root of all
individual polarization amplitudes added quadratically.

\begin{figure}[!t]
\begin{center}
$\begin{array}{lr}
\includegraphics[width=0.48\textwidth]{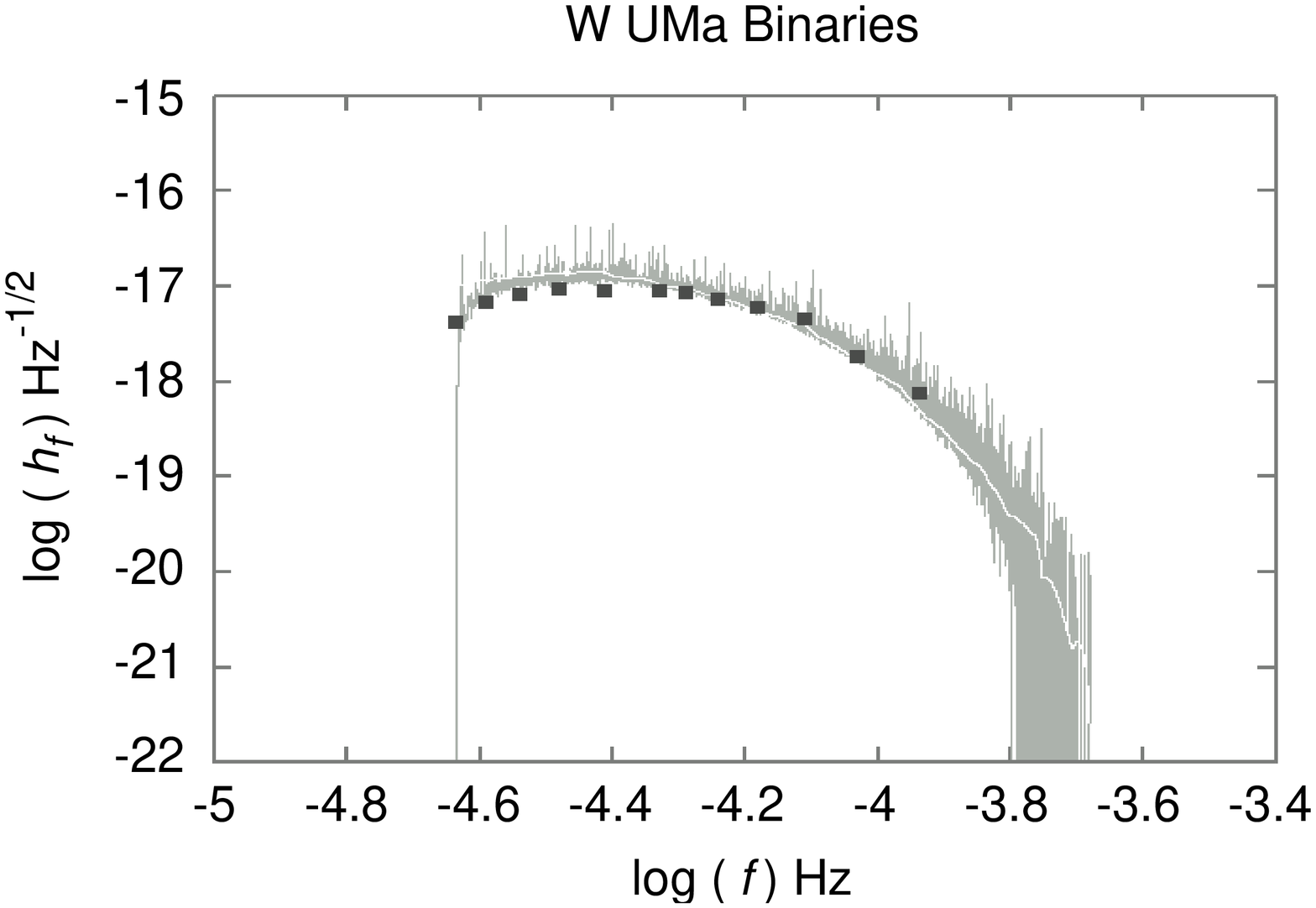}
\ & \
\includegraphics[width=0.48\textwidth]{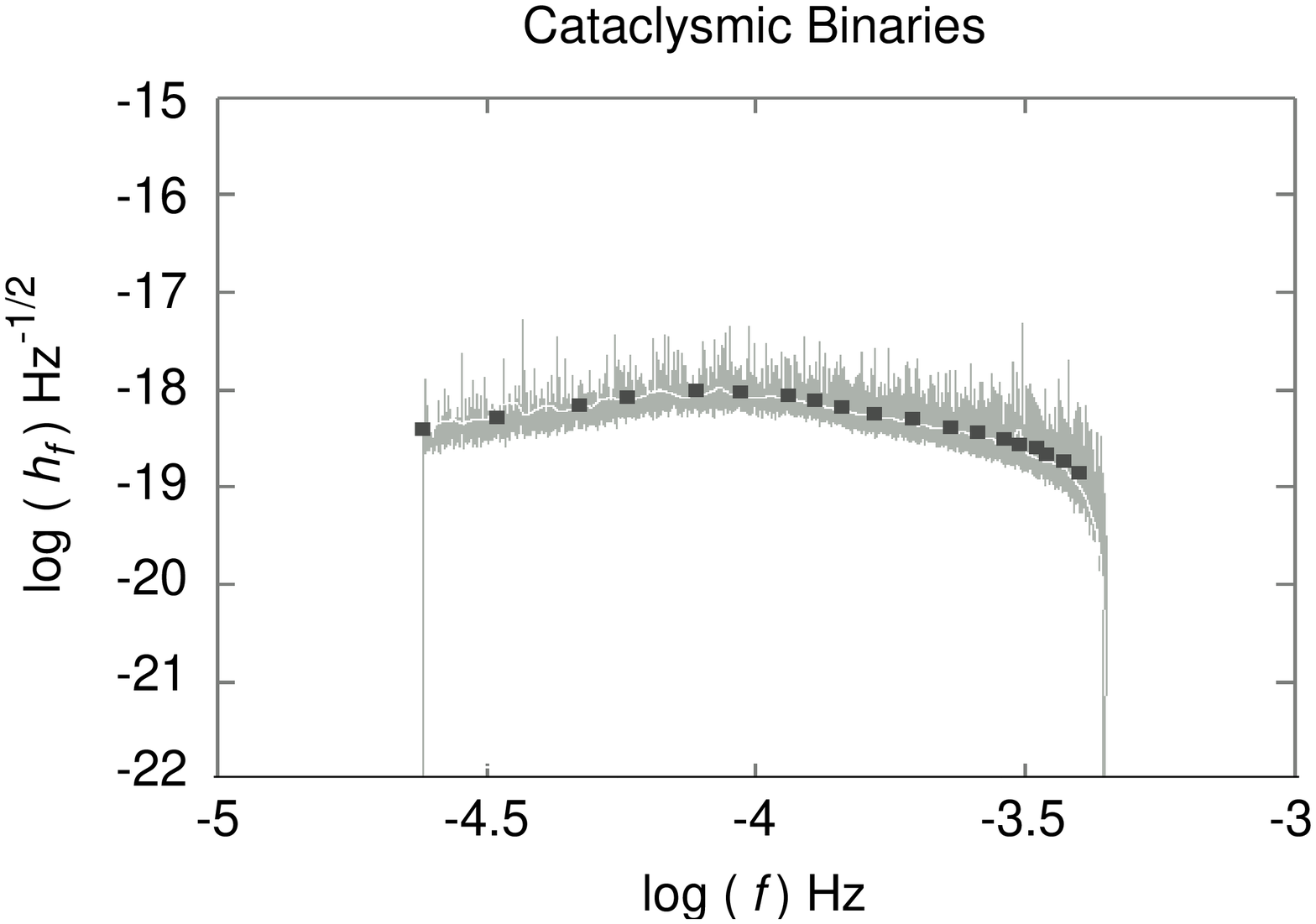}
\\
\includegraphics[width=0.48\textwidth]{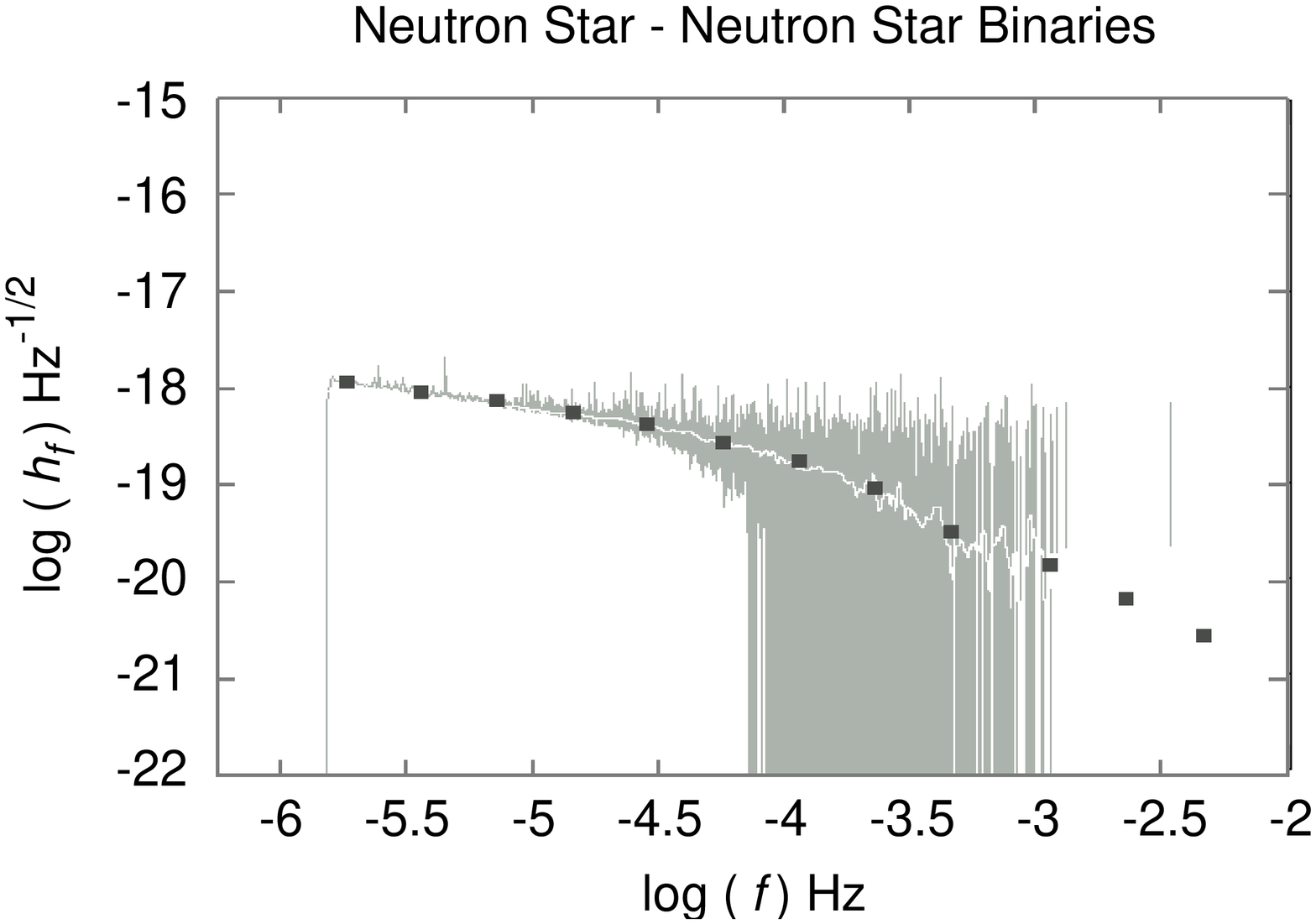}
\ & \
\includegraphics[width=0.48\textwidth]{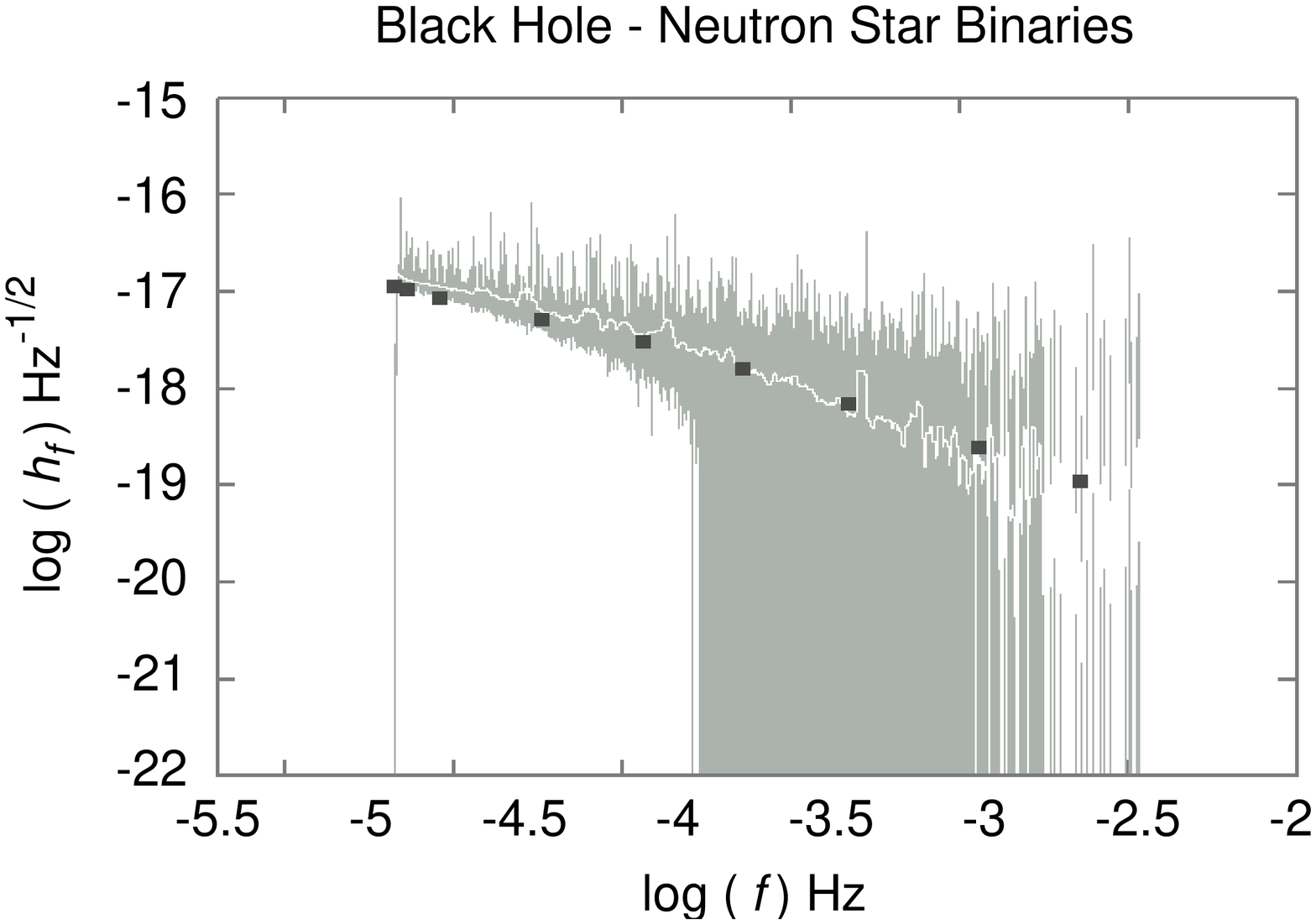}
\\
\includegraphics[width=0.48\textwidth]{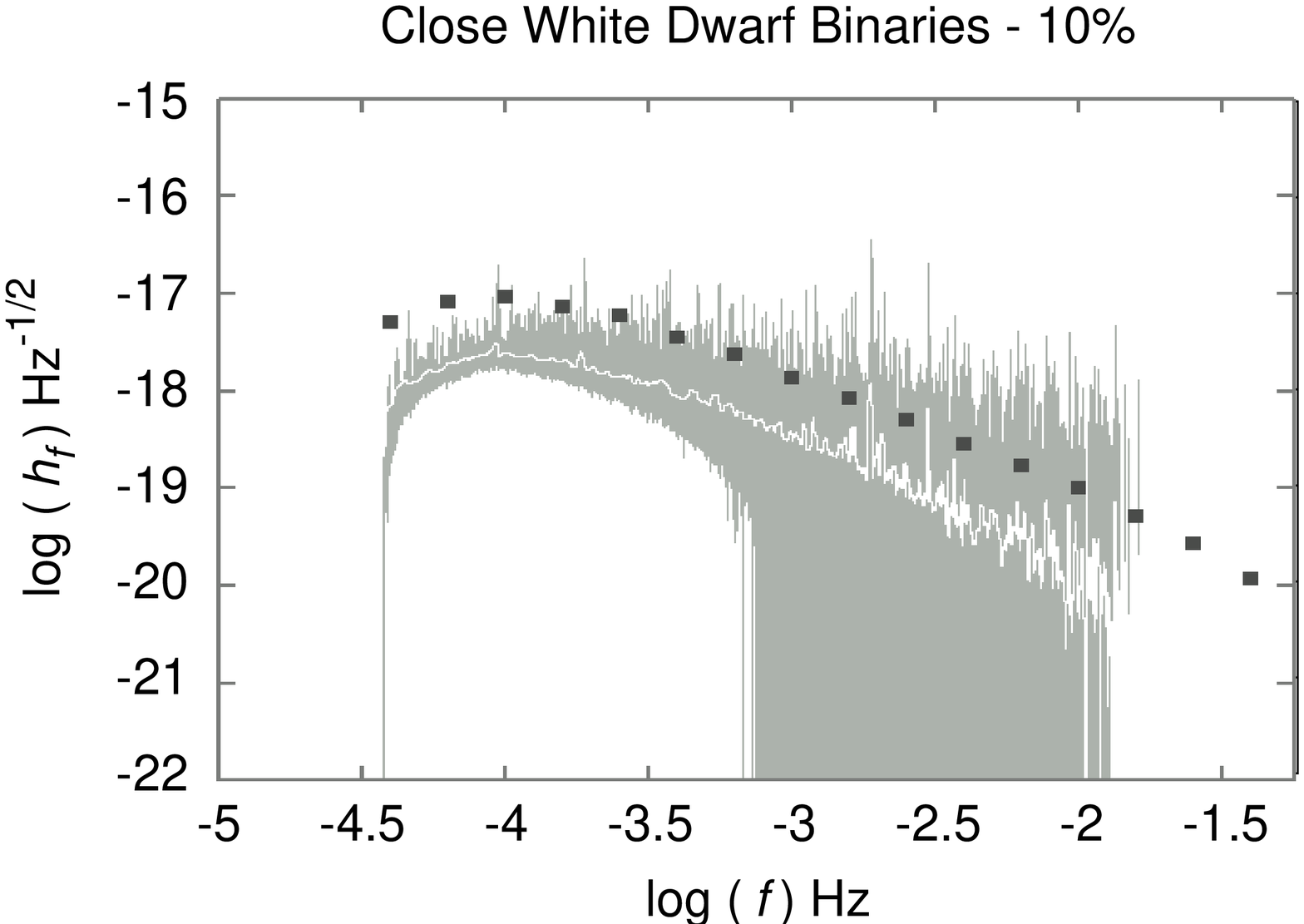}
\ & \
\includegraphics[width=0.48\textwidth]{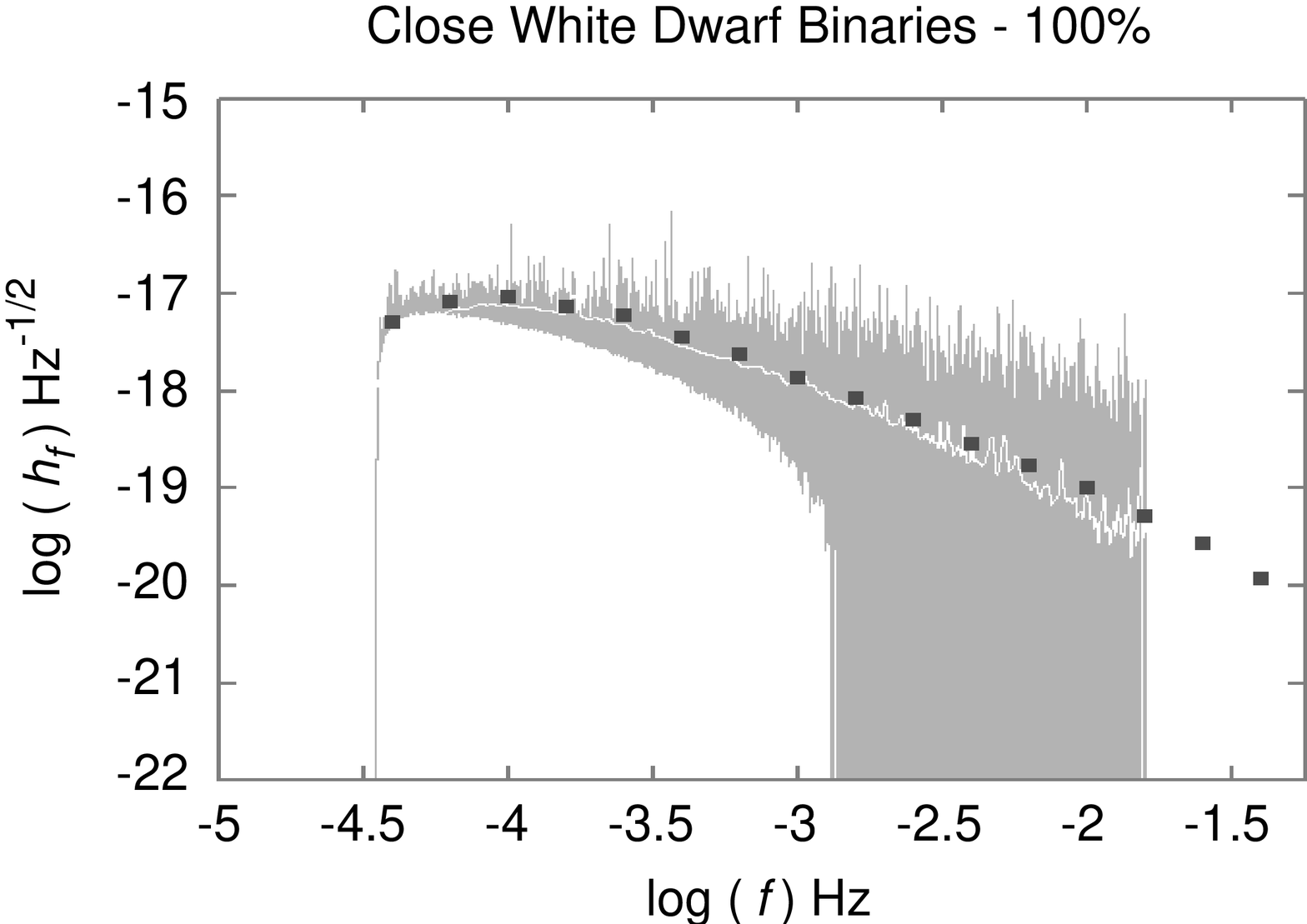}
\end{array}$
\end{center}
\caption{A comparison of our full Monte Carlo simulation (gray) and a
  running average of the simulation (white) to those of HBW
  (squares). For each binary type we are in agreement.  To ease
  comparison, the final figure shows a background using the full
  realization of white dwarf binaries.  However, our simulated
  background uses the 10\% reduce white dwarf population.}
\label{fig:sources_barycenter}
\end{figure}

Figure~\ref{fig:sources_barycenter} compares our Monte Carlo results
to HBW.  (Note that the plots show spectral amplitudes $h_f$, not
strain amplitudes $h$.  For monochromatic binaries the two are related
by $h_{f} = \sqrt{T} \, h$ where $T$ is the observational period.)
For each binary class we are in agreement.  The smearing effect seen
at high frequencies is due to empty bins in the spectrum.  For the
compact binaries the orbital period distributions have a small, but
finite, probability at short periods.  Consequently it takes a large
number of draws against the period distribution to produce a source
with an extremely short period.  In the cases of the neutron star -
neutron star and close white dwarf binaries the probability in the
period distribution tails becomes small enough that for the number of
sources included in our simulated background one would not expected a
large number of extremely short period (high frequency) sources.  This
is why the HBW data extends beyond our simulated backgrounds.

\begin{figure}[!t]
\begin{center}
\includegraphics[width=0.80\textwidth]{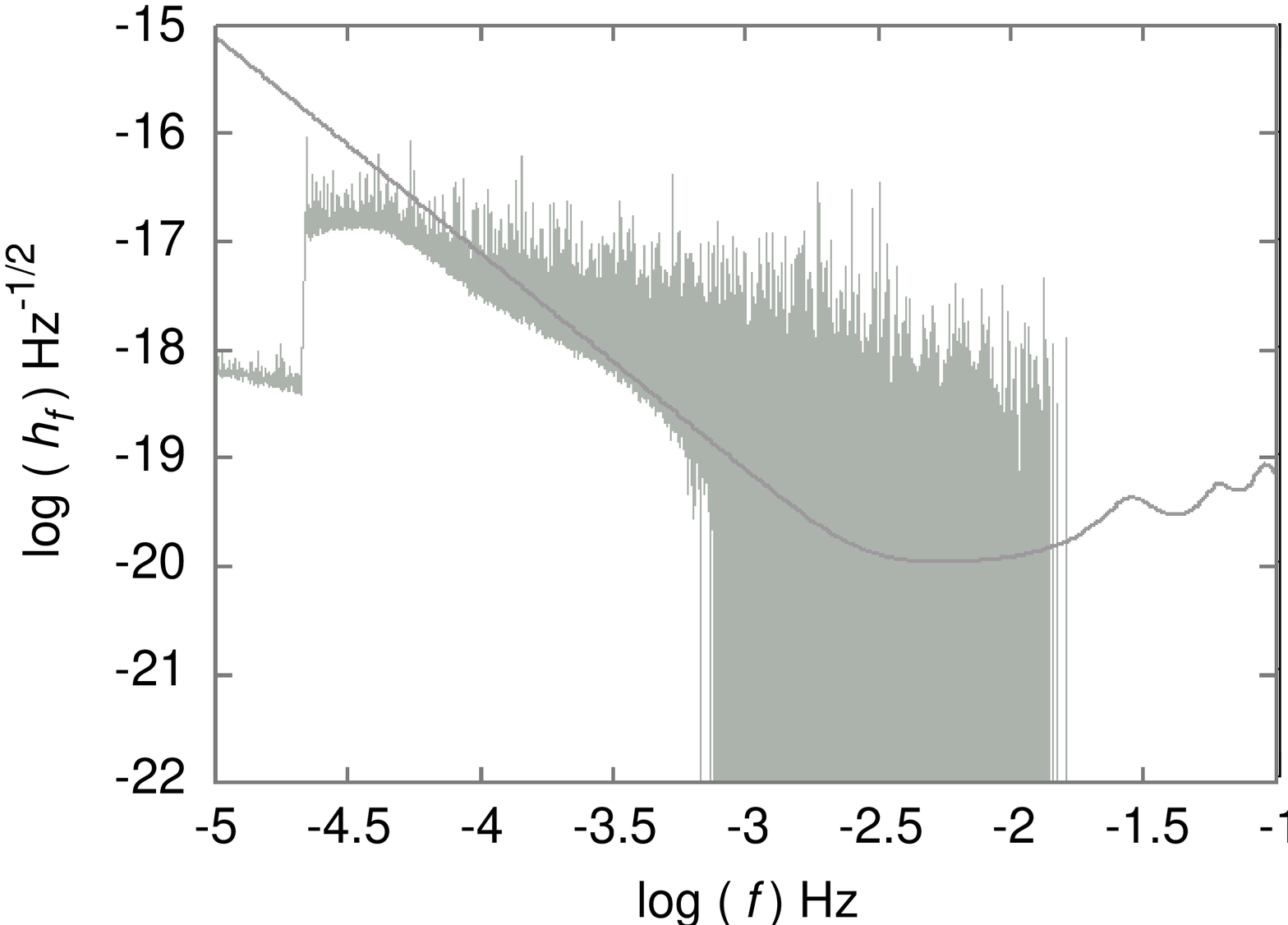}
\end{center}
\caption{A realization of the galactic background as observed in the
  barycenter frame.  The dark line is the all sky and polarization
  averaged LISA sensitivity curve \citep{LHH00}.  The jump at
  $10^{-4.6}$~Hz is due to sudden increase in the number of W~UMa
  binaries.  Had we included a realization of the unevolved binaries,
  the background levels would be roughly constant below
  $10^{-4.6}$~Hz with a spectral amplitude of $h_f \approx
  10^{-17}~\textrm{Hz}^{-1/2}$.}
\label{fig:barycenter_background}
\end{figure}

Figure~\ref{fig:barycenter_background} shows the total galactic
background as viewed in the barycenter frame.
The sharp rise in the background at $f~=~10^{-4.6}$~Hz is due to the
sudden increase in the number of W~UMa binaries. The signal below
$f~=~10^{-4.6}$~Hz is due to neutron star - neutron star binaries.  If
the unevolved binaries had been included, the galactic background
would be approximately constant between 1 and 100~$\mu$Hz at a level
of $h_f \approx 10^{-17}~\textrm{Hz}^{-1/2}$.

As \citet{NYZ01} found with their population synthesis models of the
galaxy, when the individual sources are modeled the background appears
spiky.  The large fluctuations in the background are due to a small
number of bright sources.  As we will show later, when these sources
are removed from the background the spectrum becomes smooth.

\begin{figure}[!t]
\begin{center}
\includegraphics[width=0.80\textwidth]{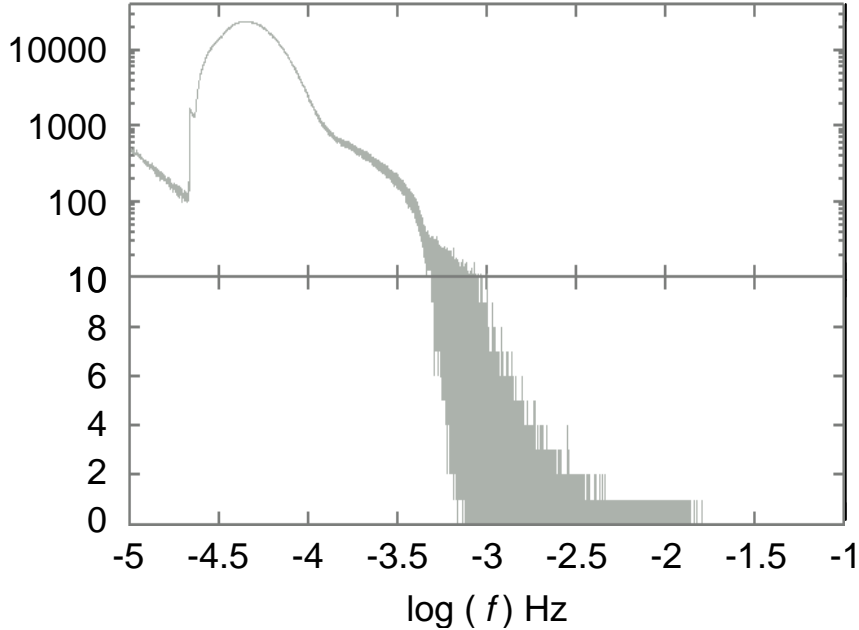}
\end{center}
\caption{Number of sources per frequency bin.  Again the dip in the
  number of sources at $10^{-4.6}$~Hz is associated with the neglected
  unevolved binaries.}
\label{fig:num_per_bin}
\end{figure}

It is generally agreed that a confusion limited background arises when
the average number of
source per frequency bin is larger than unity. (Additionally the net
strain per bin must also be larger than the intrinsic detector noise.)
Figure~\ref{fig:num_per_bin} shows that the peak number of sources per
bin is in excess of $10^{5}$ near 0.04~mHz.
The plot also demonstrates that for a large portion of LISA's spectrum
the number of sources per bin is greater than ten.

\section{DETECTOR BACKGROUND} \label{sec:detector_background}

The backgrounds presented in the previous section were not convolved
with a model for the instrument response. As with all fields of
astronomy, the act of measuring incident radiation has to be properly
understood in order to correctly interpret the signals. In this
section we first summarize the LISA mission and then describe our
models for the detector response.  We then present the simulated
galactic background as measured by the detector.

LISA is a joint ESA/NASA mission planned for launch around 2015.  The
mission consists of three identical spacecraft in separate, slightly
eccentric, heliocentric orbits inclined with respect to the ecliptic
plane \citep{DR03}.  The orbits are carefully chosen such that the
spacecraft constellation will form, and approximately maintain, an
equilateral triangle with a mean spacecraft separation of
$5~\times~10^{6}$~km. The center of the constellation, referred to as
the guiding center, will have an orbital radius of 1~AU and trail the
Earth by $20^{\circ}$.  During the course of one orbit, the
constellation will cartwheel once with a retrograde motion as seen by
an observer at the Sun.  LISA is sensitivity to gravitational
radiation in the range of $10^{-5}$ to 1~Hz.

The detector's motion introduces amplitude (AM), frequency (FM), and
phase modulations (PM) into the gravitational wave signals
\citep{CL03b}.  The amplitude modulation originates from the
detector's motion sweeping the antenna pattern across the sky.  The
phase modulation results from the differing responses to each
polarization state.  The frequency (Doppler) modulation is due to the
motion of the detector relative to the source.  Since the bulk orbital
and cartwheel motions both have a period of one year, the resulting
modulations appear as sidebands in the power spectrum separated from
the instantaneous carrier frequency by integer values of the
modulation frequency, $f_{m} = 1/\textrm{year}$.

For our studies the detector response is modeled using a combination
of the \textit{Extended Low Frequency Approximation}, which is
developed in the appendix, and the \textit{Rigid Adiabatic
Approximation} as described in \citet{RCP04}.  As explained in the
appendix, to save on computational costs it is advantageous to
simulate the detector response directly in the frequency domain where
a quasi-monochromatic source will only have a small number of
sidebands. At low frequencies, the bandwidth for a slowly evolving
circular binary is
\begin{equation}
  B = 2 \left( 4 + \frac{2\pi f R}{c} \sin (\theta) \right) f_m \,,
\end{equation}
where $R =1$~AU is the orbital radius of LISA and $\theta$ is the
colatitude of the source on the celestial sphere.  For sources with
gravitational wave frequencies below a few millihertz, the bandwidth
is less than $100 f_{m}$ and we can achieve a considerable saving in
computational cost by working in the frequency domain. At higher
frequencies the sources have larger bandwidths, more complex
modulation patterns, and signals that evolve significantly in
frequency, making them harder to model directly in the frequency
domain.

\begin{figure}[!t]
\begin{center}
\includegraphics[width=0.80\textwidth]{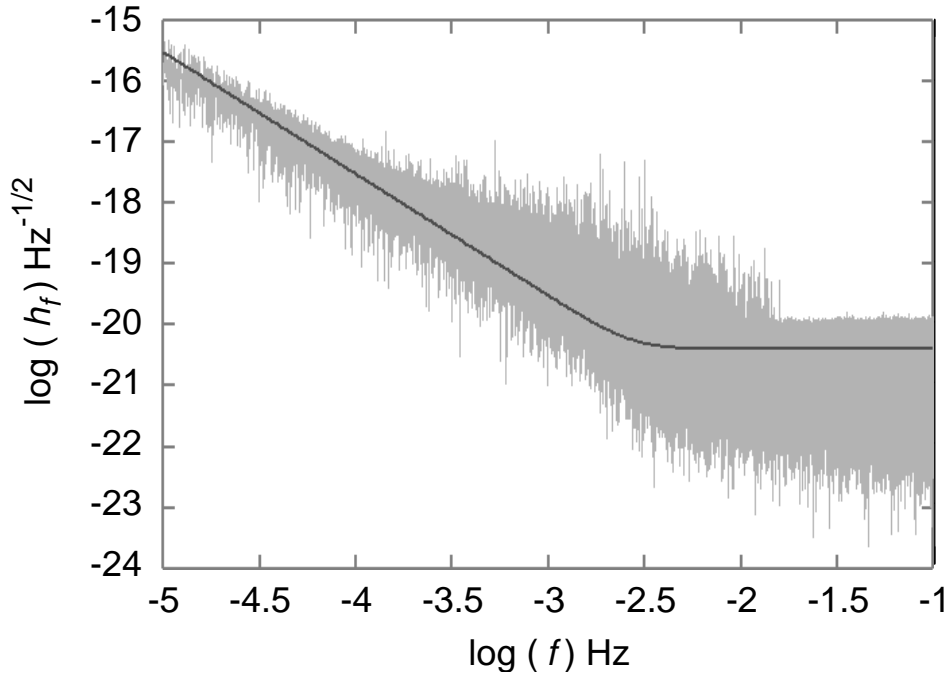}
\end{center}
\caption{A realization of the HBW 10\% galactic background as measured in the
  detector's frame.  The dark line is the average Michelson noise
  associated with the detector.  The galactic gravitational wave
  background is evident in the spiky structure between 0.1 and
  10~mHz.}
\label{fig:detector_background}
\end{figure}

\begin{figure}[!t]
\begin{center}
\vspace{2cm}
\includegraphics[width=0.80\textwidth]{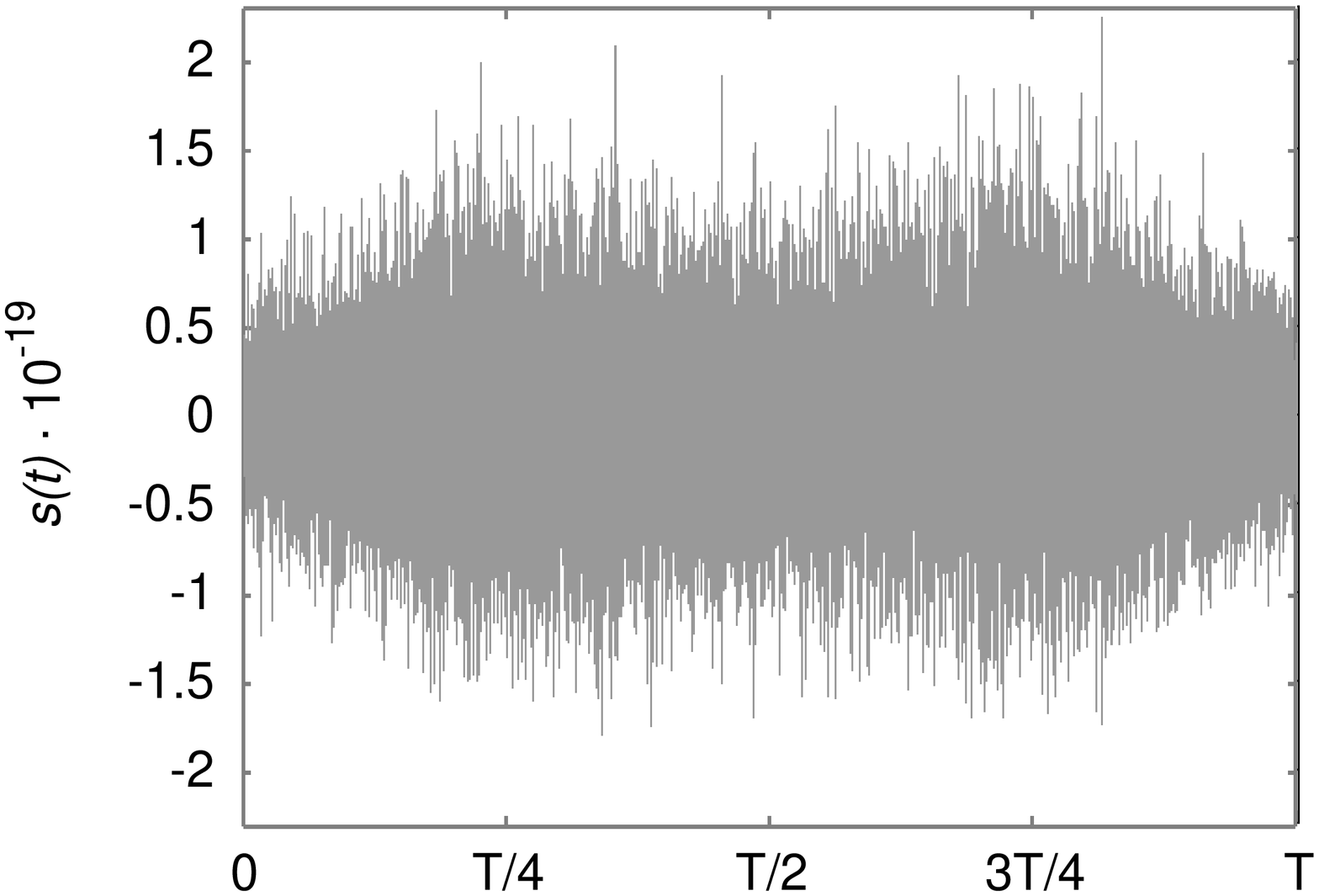}
\end{center}
\caption{A realization of the HBW 10\% galactic background as measured in the
  detector's frame shown in the time domain (with no instrument
  noise).  The annual modulation of the signal due to the detector
  motion is apparent by the two large lobes.}
\label{fig:detector_background_time}
\end{figure}

In the appendix it is shown that the \textit{Extended Low Frequency
Approximation}, which is applied in the frequency domain, is only
valid for frequencies below 7~mHz.  For the few hundred signals with a
carrier frequency above this cutoff, we use the more accurate time
domain response model, the \textit{Rigid Adiabatic Approximation}.  To
combine the results from each approximation, we first simulate the
response for the signals that are above 7~mHz using the more detailed
\textit{Rigid Adiabatic Approximation} and add them linearly in the
time domain.  We then perform a Fast Fourier Transform.  For the
sources that are simulated directly in the frequency domain, we
coherently add the signals by summing the real and imaginary parts of
their respective Fourier coefficients at each frequency.  By adding
the coefficients we maintain the phase information which dictates the
constructive and destructive interference of the gravitational waves.
The final detector response is the sum of the Fourier coefficients
from the extended low frequency and adiabatic results.  It is also at
this time that we add in a detector noise realization using the
prescription given in \citet{RCP04}.
Figure~\ref{fig:detector_background} shows a particular realization
for a Michelson signal in the frequency domain.
Figure~\ref{fig:detector_background_time} shows the same signal in the
time domain.

In comparing the background as observed in the detector frame
(fig.~\ref{fig:detector_background}) versus in the barycentric frame
(fig.~\ref{fig:barycenter_background}) a striking feature is that it
is lower by roughly a full decade across the entire spectrum.  The
reduction is due to two effects. The first effect is the detector
efficiency, which relates the total signal power in the detector to
the total signal power at the barycenter.  The all-sky and
polarization averaged detector efficiency is equal to $\sqrt{3/20}$ at
low frequencies, and gets progressively worse at high frequencies.
(This is immediately evident by comparing the sensitivity curve in
figure~\ref{fig:barycenter_background} to the average Michelson noise
in figure~\ref{fig:detector_background}.)  The second effect is due to
the orbital motion of the detector.  In the barycentric frame, most
galactic binaries are well approximated as monochromatic.  As LISA
moves in its orbit the monochromatic signals are modulated across
multiple frequency bins.  At high frequencies the spreading effects
are evident by spectral power showing up in bins that were previously
empty in the barycenter frame.  At low frequencies the expectation is
that the spreading from adjacent bins will cancel out due to the
numerous sources found in each bin (see fig.~\ref{fig:num_per_bin}).
However, as will be shown in the next two sections, the galactic
background is dominated by a few bright sources.  When the bright
sources are modulated there is not a compensating bright source in the
adjacent bin.  As a result the galactic background is also reduced at
lower frequencies.

\section{STATISTICAL CHARACTER OF THE GALACTIC BACKGROUND} \label{sec:gauss}

Of great interest to the LISA mission is to broadly characterize the
galactic gravitational background in a statistical sense.  Such a
characterization is essential to the development and implementation
of data analysis algorithms which often make assumptions about the
character of the noise.

In the spectral regions of LISA's band where the galactic background
dominates the detector response, the background becomes a source of
noise.  By inspection of the spectrum in
figure~\ref{fig:detector_background}, the galaxy is evident by the
jaggedness between 0.1 and 10~mHz.  Outside this region the galactic
binary signals are weaker than the intrinsic detector noise.  This is
evident in the plot by the relative smoothness of the spectrum from
bin-to-bin.  One way to characterize the background is to
statistically study the Fourier coefficient distributions in different
regions of the spectrum.  Of specific interest is finding out if the
galactic background is characterized by a Gaussian distribution.

\begin{figure}[!t]
\begin{center}
\includegraphics[width=0.80\textwidth]{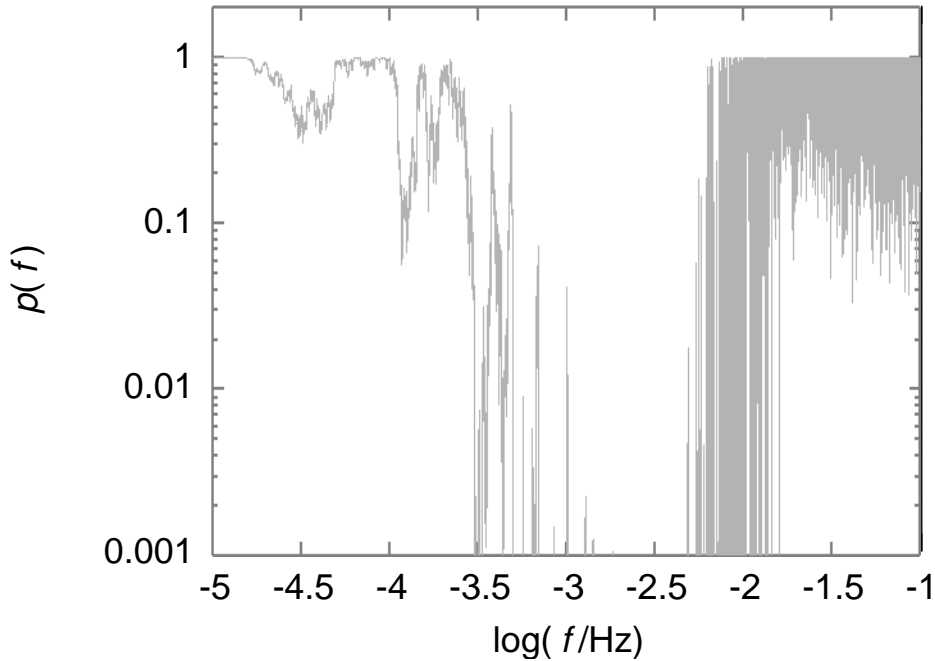}
\end{center}
\caption{The Kolmogorov-Smirnov Gaussianity test applied to the real Fourier
  coefficients for the detector output shown in
  figure~\ref{fig:detector_background}.  The presence of the galactic
  background is apparent by the low $p$ values about $f = 10^{-3}$~Hz.
  Similar results are found for the imaginary coefficients, for the
  other data channels, and when using a $\chi^2$ test.}
\label{fig:gauss_test}
\end{figure}

Tests for Gaussianity are done using independent $\chi^2$ and
Kolmogorov-Smirnov tests.  The Gauss tests are performed over a window
of 512 bins and done at each frequency.  Figure~\ref{fig:gauss_test}
shows the results of the Kolmogorov-Smirnov test performed on the real
coefficients for the spectrum shown in
figure~\ref{fig:detector_background}.
The $p$ values plotted along the ordinate axis are the probabilities
that the set of measured deviates are Gaussian distributed.  While it
is impossible to say with certainty that a set of measured deviates
are necessarily distributed with a specific distribution, they can be
shown not to agree.  Values for $p$ above a hundredth are usually
accepted as agreeing with the tested distribution function.  For $p$
values below a hundredth, which are seen in the millihertz region of
the detector output, indicate that the measured set of data is not
Gaussian distributed.

Comparing the results of the Gaussian test (fig.~\ref{fig:gauss_test})
and the original LISA output (fig.~\ref{fig:detector_background})
indicates that the detector response is non-Gaussian for frequencies
in which the background is above the intrinsic detector noise level.
Outside these regions, where the detector's noise dominates the
output, the returned $p$ values are consistent with a Gaussian
distribution, as they should since the simulation of the noise is
based on Gaussian distributions.  Similar results are found for the
imaginary coefficients, other channels of data, and when calculated
using a $\chi^2$ test.

A common misconception is that the galactic background should be
Gaussian distributed.  This assumption is based on the Central Limit
Theorem, which states that for a large sample of random deviates,
regardless of their parent distribution, the distribution of average
values will be approximately Gaussian.  However, the Central Limit
Theorem is not directly applicable to the galactic background since
the net power in a single frequency bin may be dictated by a single
bright source.  Moreover, in a small region of the spectrum there are
very few bright sources.  Far less than the large number of data
points required for the Central Limit Theorem to apply.

\section{CONFUSION LIMITED BACKGROUND} \label{sec:clb}

\subsection{Gaussian Nature of the Confusion Limited Background}

Large fluctuations originating from bright sources cause the tails of
the expected Gaussian distributions to be enlarged.  If the bright
sources were removed from the data streams, then the remaining
background would be Gaussian.  To demonstrate this we first identify
the bright sources, subtract them, and then retest the remaining
background for Gaussianity.

Identifying all the bright sources in the actual LISA data streams is a
difficult problem not yet fully solved.  The modulation effects caused
by LISA's orbital motion spread a source's spectral power across
multiple frequency bins.  Although the bandwidth over which a signal
will spread is a known function of the gravitational wave frequency
and sky position, if multiple signals are overlapping in a small
region of frequency space, the true number of signals in the region
may not be clearly identifiable.  For our Monte Carlo models, when we
generate each binary, the parameter values for each system are known.
With this extra information we can quickly and accurately identify
bright sources and remove them from the data streams.

Our approach to identifying bright sources is to categorize them
according to their signal-to-noise (SNR) ratio using the standard
formula,
\begin{equation} \label{eq:snr}
  ({\rm SNR})^2 = 4 \int_{0}^{\infty} \frac{ | \widetilde{h}(f)
  |^2}{S_{n}(f)} \, df \,,
\end{equation}
where $\widetilde{h}(f)$ is the Fourier transform of the noiseless
response to a single gravitational wave signal, and $S_{n}(f)$ is the
one-sided noise power spectral density.  A source is labeled as
``bright'' if its SNR is greater than 5 (optimistic) or
10 (conservative).  The proper use of
equation~\eqref{eq:snr} requires a clear interpretation of the noise.
We are interested in removing sources that are bright relative to the
local power spectrum level.  Therefore, the $S_{n}(f)$ curve must be a
composite of the intrinsic detector noise and the galactic background.
It is an effective noise for the detector, which we will emphasize by
denoting it as $S_{n}^{\textrm{eff}}(f)$.

To approximate the effective noise we calculate the median detector
output.  While representing the effective noise by the median response
(as opposed to the mean response) is somewhat stable against bright
sources, in regions near an extremely large SNR signal or where the
density of bright sources is large, the median will still be
influenced by these few signals.  To account for the influence of the
bright sources in calculating the $S_{n}^{\textrm{eff}}(f)$ curve, we
perform our calculations iteratively.  We start with the median
response of the initial output and calculate the number of bright
sources with reference to this curve.  We then remove the bright
sources exactly using the same detector response approximation that
generated them.  The justification for using the exact parameters is a
matter of simplicity, since data analysis algorithms are still being
developed. Moreover, with a SNR threshold of 10 the errors in the
recovered parameters will be small. The result of removing the bright
sources is a new background from which we can calculate a new median
response.  From the new median response we calculate the number of
bright sources with respect to the new $S_{n}^{\textrm{eff}}(f)$
curve.  We iterate this procedure until there are no new bright sources
being identified. In some examples, such as with the 100\% HBW
model, the procedure does not appear to converge, so the subtraction
was stopped after 10 iterations. 

Previous estimates~\citep{BH97,BC04} of the confusion background ignored
the relative brightness of the sources and focused instead on the source
density. These estimates defined the confusion regime in terms of the
number of sources per frequency bin. Outside of the confusion regime
sources could be resolved and removed, while inside the confusion
regime the sources acted as a source of noise. This notion of source
confusion is based on linear algebra: The signal from each galactic binary
is described by 7 or 8 parameters~\citep{CP05}, and there are 4 data
points per frequency
bin (two independent channels, each with a real and imaginary part). Thus,
one needs at least 2 frequency bins per source to have as many data points
as there are unknowns. While this estimate is very crude, it is unlikely
that a data analysis algorithm can be found that beats the 0.5 source
per bin saturation point by very much (methods such as Maximum
Entropy~\citep{jaynes} introduce priors that can help tame under
constrained systems, but they can only do so much). In order to study
the effect of a source density cut-off we repeat the SNR based
source subtraction procedure with a maximum resolved source density
of 0.25 sources per bin using a 100 bin average. The density of
one source per four bins was chosen as intermediate between the
capabilities of existing algorithms~\citep{CC05} and the saturation
point described above.

%Table~\ref{tab:outliers}
Table 1 lists the number of bright sources
removed at each iteration and the total number of bright sources
removed for several different realizations of the galactic
background if no source density cut-off is applied. In each of
the three HBW 10\% realizations, the total number of
sources was fixed at $3.635~\times~10^{7}$.  All three realizations
gave similar results.  The results of a HBW background using the full
100\% of the white dwarves, along with the NYZ white dwarf
background ($2.6~\times~10^7$ sources) are also shown.
For comparison we also include the optimistic case of using a
subtraction threshold of ${\rm SNR=5}$. 

\begin{deluxetable}{ccccccc}\label{tab:outliers}
\tablewidth{0pt}
\tablecolumns{8}
\tablecaption{Number of New Bright Sources Identified at Each Iteration}
\tablehead{
\colhead{Iteration} &
\colhead{10\%(1)} &
\colhead{10\%(2)} &
\colhead{10\%(3)} &
\colhead{100\% } &
\colhead{10\% (SNR=5)} &
\colhead{NYZ}}
\startdata
1 & 6795 & 6848 & 6793 & 10,736 & 14,346 & 8583 \\
2 & 3806 & 3723 & 3712 & 8084   & 10,620 & 4803 \\
3 & 1693 & 1803 & 1846 & 5340  &    5480 & 2090 \\
4 & 817 & 929   & 866  & 3550  &    2577 & 866 \\
5 & 395 & 457   & 456  & 2597  &    1129 & 473 \\
6 & 226 & 250   & 225  & 2014  &     525 & 266 \\
7 & 114 & 172   & 123  & 1628  &     278 & 191  \\
8 &  59 &  97   &  76  & 1400  &     144 & 125  \\
9 &  -  &  -    &  -   & 1264  &      72 &  -   \\
10&  -  &  -    &  -   & 1136  &      38 &  -   \\
\hline
Total & 13,905 & 14,279 & 14,097 &  37,749 & 35,209  & 17,396\\
\enddata
\end{deluxetable}

\begin{figure}[!t]
\begin{center}
\vspace{2cm}
\includegraphics[width=0.80\textwidth]{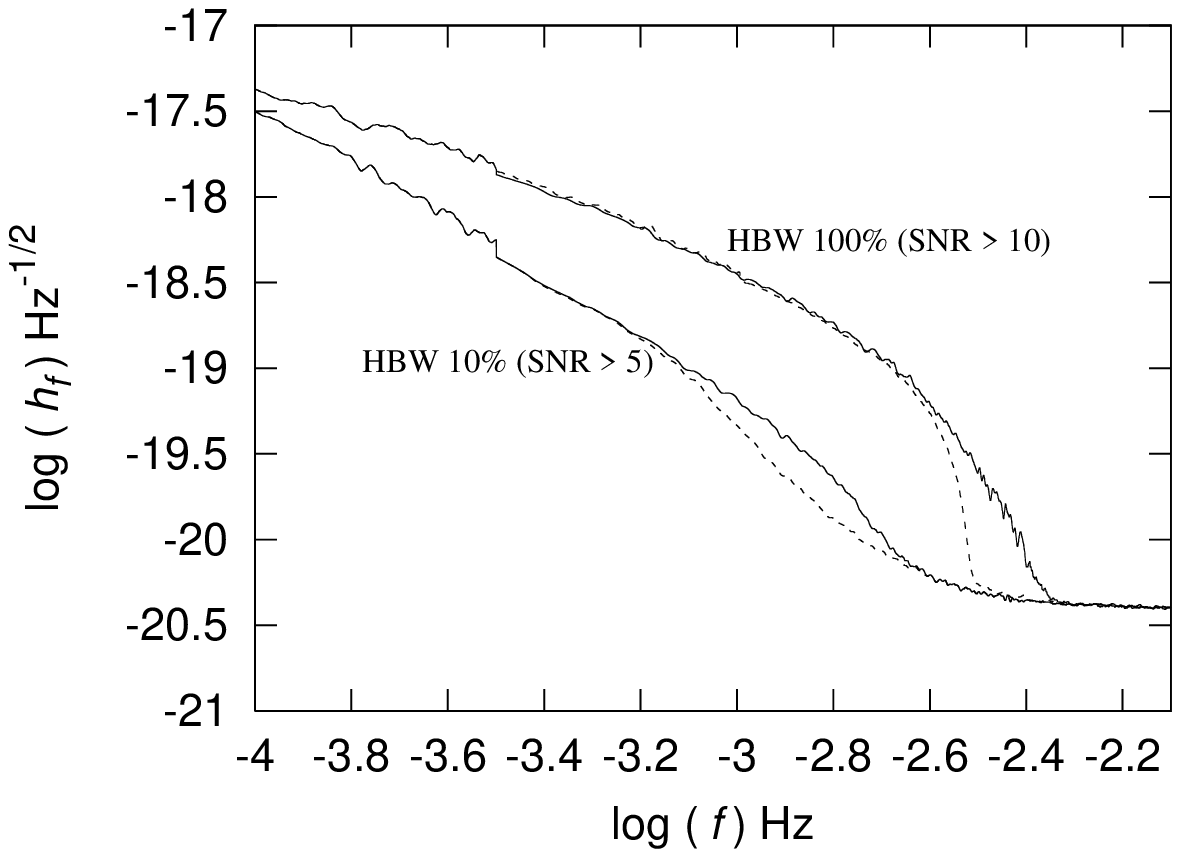}
\end{center}
\caption{Effective noise levels for the the HBW 100\% (${\rm SNR} > 10$)
and HBW 10\% (${\rm SNR} > 5$) models with (solid) and without (dotted)
a source density cut-off of one resolved source per four frequency
bins.}
\label{fig:cut_vs_nocut}
\end{figure}

%Table~\ref{tab:cut}
Table 2 lists the number of bright sources
removed at each iteration and the total number of bright sources
removed for several different realizations of the galactic
background when a maximum density of 0.25 resolved sources
per bin is applied. The HBW 10\% and NYZ models are only slightly
affected by the cut-off, while the HBW 100\% and ${\rm SNR} > 5$ version
of the HBW 10\% model are significantly affected. The impact
of the cut-off is evident in Figure~\ref{fig:cut_vs_nocut}, where
the effective noise levels for the HBW 100\% (${\rm SNR} > 10$)
and HBW 10\% (${\rm SNR} > 5$) models are shown with and without the
source density cut-off applied.

\begin{deluxetable}{ccccc}\label{tab:cut}
\tablewidth{0pt}
\tablecolumns{6}
\tablecaption{Number of New Bright Sources Identified at Each Iteration with
a Source Density Cut-Off Applied}
\tablehead{
\colhead{Iteration} &
\colhead{10\%} &
\colhead{100\% } &
\colhead{10\% (SNR=5)} &
\colhead{NYZ}}
\startdata
1 & 6795  & 10,736 & 14,346 & 8583 \\
2 & 3751  & 7850   &  4898  & 4803 \\
3 & 1669  & 4440   &   194   & 2007 \\
4 & 724   & 1531   &    20    & 732 \\
5 & 271   & 463    &    1     & 325 \\
6 & 79    & 157    &    -     & 186 \\
7 & 30    &  66    &    -     & 79  \\
8 & 12    &   1     &    -     & 48  \\
9 &  7    &   -    &    -     & 27  \\
\hline
Total & 13,338  &  25,244 & 19,459 & 16,788\\
\enddata
\end{deluxetable}

\begin{figure}[!t]
\begin{center}
\includegraphics[width=0.80\textwidth]{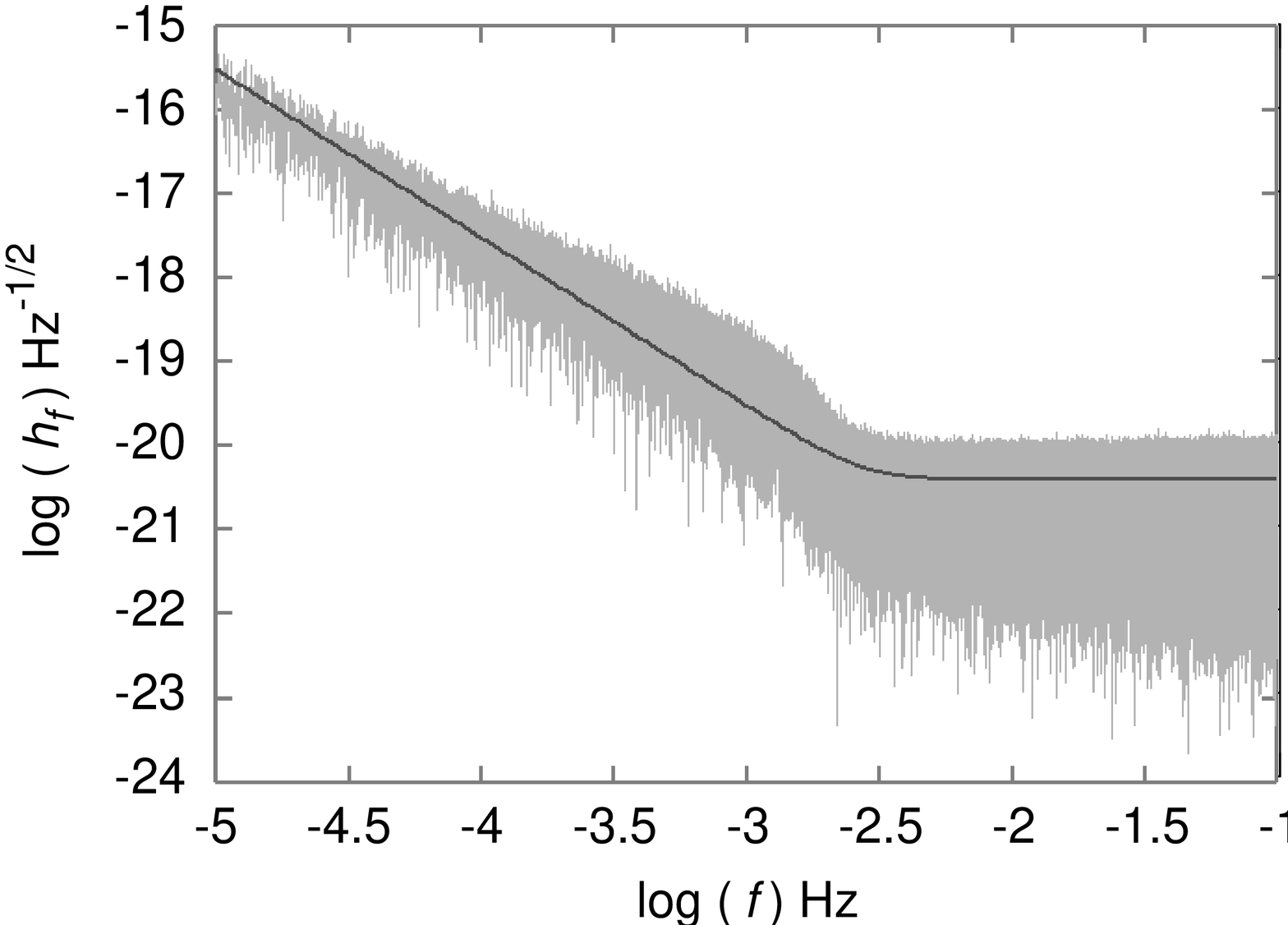}
\end{center}
\caption{The same Michelson signal from before, but after the
  $\sim$$10^{4}$ bright sources have been removed from the data
  streams.}
\label{fig:background_rem}
\end{figure}

\begin{figure}[!t]
\begin{center}
\includegraphics[width=0.80\textwidth]{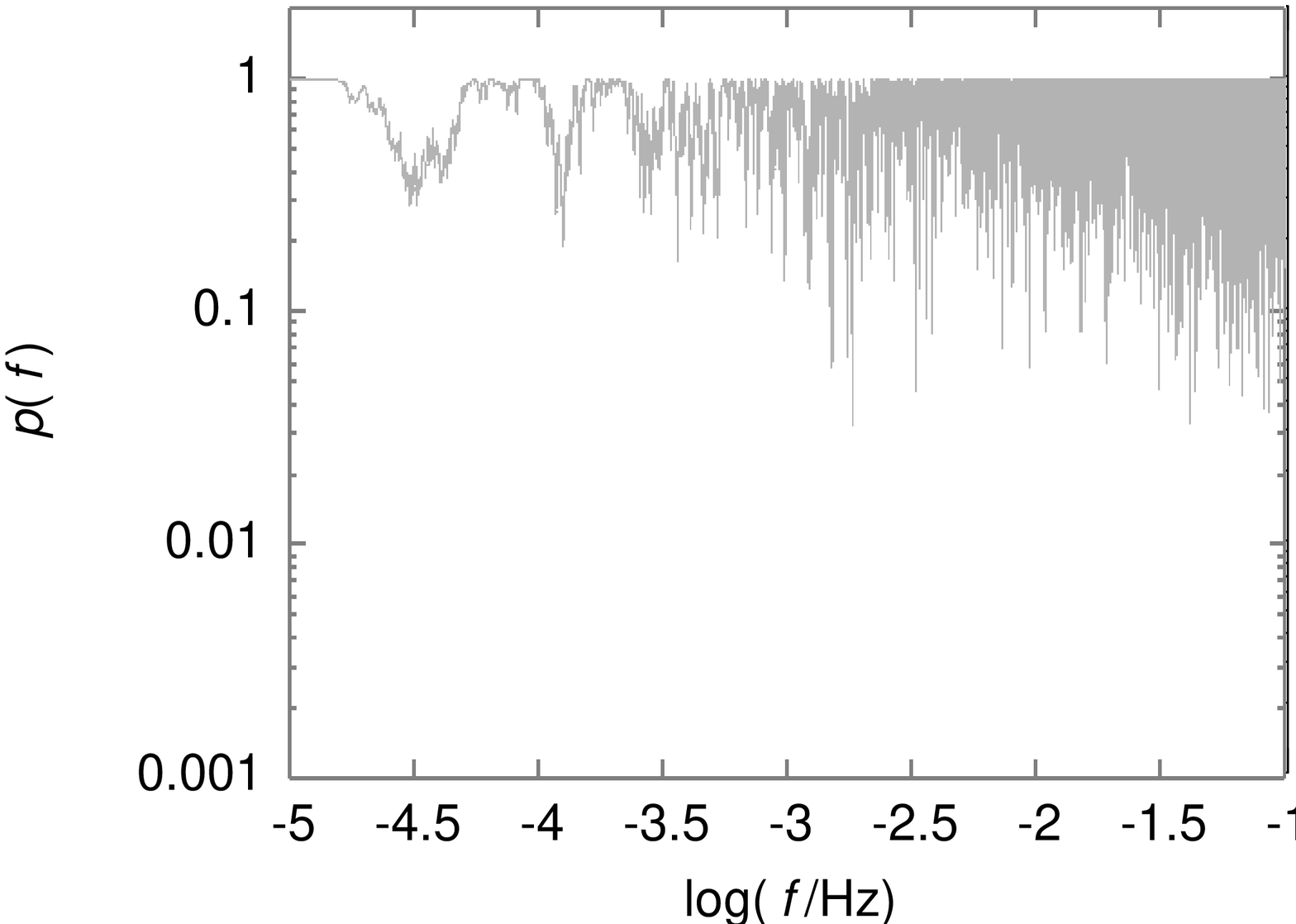}
\end{center}
\caption{A running Gauss test applied to the detector output after the
  bright sources have been removed.  Unlike before, all returned $p$ 
  values are consistent with a Gaussian distribution.}
\label{fig:gauss_test_rem}
\end{figure}

Shown in figure~\ref{fig:background_rem} is the same Michelson signal
as before (fig.~\ref{fig:detector_background}), but with the the
bright sources removed.
Visual inspection of the spectrum shows that without the bright
sources the bin-to-bin fluctuations are much smaller; an indication
that the Fourier coefficients may be Gaussian distributed.
Figure~\ref{fig:gauss_test_rem} confirms this hypothesis.
When the bright sources are removed from the data streams the galactic
background is Gaussian in nature.

\subsection{Confusion Limited Background Estimate}

The sources that are not flagged as bright will give rise to a
confusion limited background that acts as an effective noise source
for LISA. Taking the list of unresolved sources that remain after the
simulated data analysis procedure described in the previous section,
we can generate estimates of the confusion noise in either the
barycentric frame or in the instrument data channels. The former is
useful for making comparisons with earlier work, while the latter is
better suited to studying the effect of the confusion background on
LISA's ability to resolve other types of gravitational wave signals.

\begin{figure}[!t]
\begin{center}
\vspace{2cm}
\includegraphics[width=0.80\textwidth]{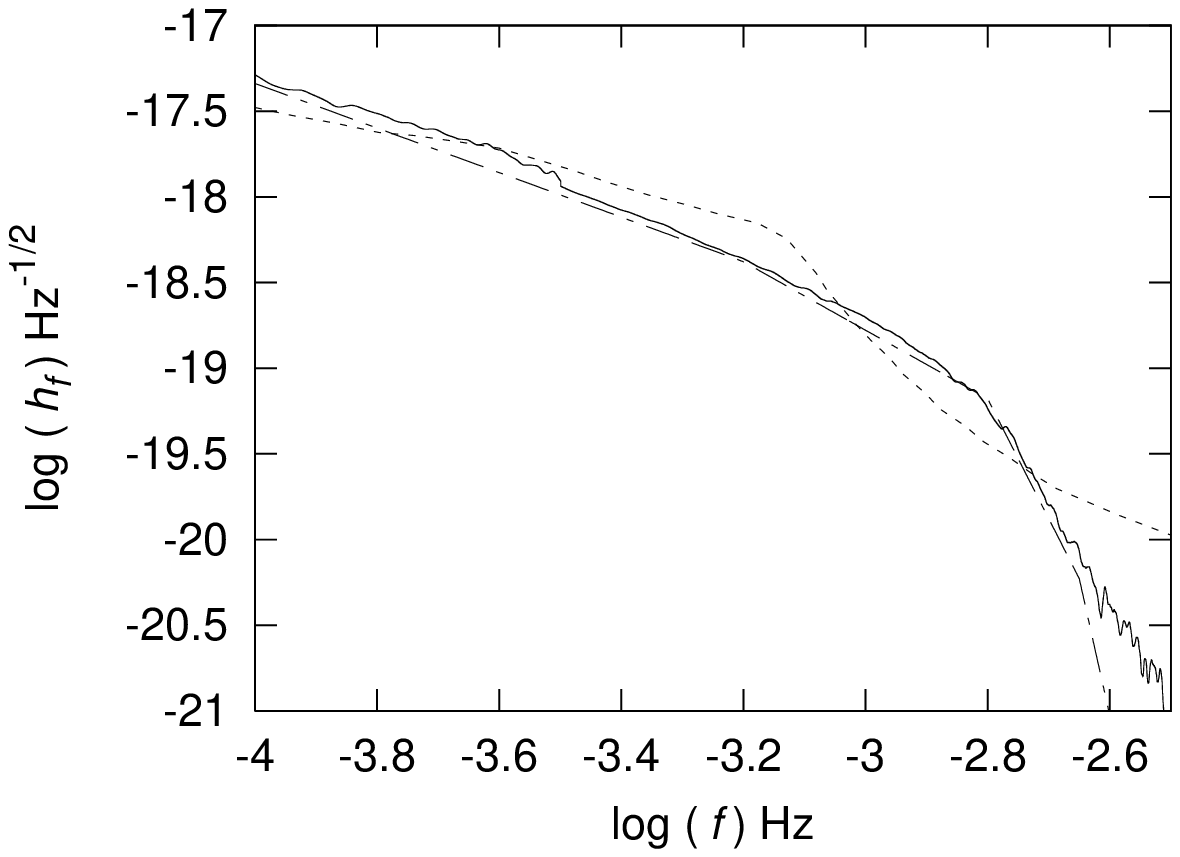}
\end{center}
\caption{Our estimates for the HBW 10\% barycenter (solid line) and detector
  frame (dot-dash line) confusion noise levels compared to the
  \citet{BH97} estimate (dotted line). Our detector frame confusion
  noise has been multiplied by a factor of $\sqrt{20/3}$ to account
  for the average detector efficiency.}
\label{fig:confusion_noise}
\end{figure}

Figure~\ref{fig:confusion_noise} compares our barycenter and detector
frame confusion noise estimates to the barycenter estimate of
\citet{BH97}. We have multiplied our detector frame result by a factor
of $\sqrt{20/3}$ to account for the average detector efficiency.  Our
estimate is lower than the \citet{BH97} estimate at low frequencies
and higher at high frequencies. It is important to note that both
estimates use exactly the same HBW model for the compact galactic
binaries. The differences in the confusion noise estimates are due to
the different way in which we modeled the data analysis
procedure. Note that our results can only be compared below
$\sim$1.3~mHz. Above this frequency the \citet{BH97} estimate is
dominated by extragalactic sources, which we did not include in our
simulation.

A simple piecewise fit to our confusion noise estimate of the HBW 10\%
background in the detector frame is given by
\begin{equation} \label{eq:clb}
  S_{\textrm{conf}}(f)  = \left\{ \begin{array}{ll}
    10^{-45.9}f^{-2.6} & \qquad 10^{-4.4} < f \le 10^{-3.2} \\
    10^{-50.38}f^{-4.0} & \qquad 10^{-3.2} < f \le 10^{-2.8} \\
    10^{-78.38}f^{-14.0} & \qquad 10^{-2.8} < f \le 10^{-2.65} \\
    10^{-126.08}f^{-32.0} & \qquad 10^{-2.65} < f \le 10^{-2.55} \\
    10^{-62.33}f^{-7.0} & \qquad 10^{-2.55} < f \le 10^{-2.1}
\end{array} \right. \; \; {\rm m}^2 \, {\rm Hz}^{-1} \,.
\end{equation}
Similarly, for the NYZ white dwarf binary background we found
\begin{equation} \label{eq:clb_n}
  S_{\textrm{conf}}(f)  = \left\{ \begin{array}{ll}
    10^{-44.62}f^{-2.3} & \qquad 10^{-4.0} < f \le 10^{-3.0} \\
    10^{-50.92}f^{-4.4} & \qquad 10^{-3.0} < f \le 10^{-2.7} \\
    10^{-62.8}f^{-8.8} & \qquad 10^{-2.7} < f \le 10^{-2.4} \\
    10^{-89.68}f^{-20.0} & \qquad 10^{-2.4} < f \le 10^{-2.0}
\end{array} \right. \; \; {\rm m}^2 \, {\rm Hz}^{-1} \,.
\end{equation}
These fits do not include instrument noise, and have been quoted in
terms of position noise in order to avoid ambiguities in the path
length scaling.  (Our simulations normalize by the round trip path
length, while other studies normalize by the detector arm length.) For
comparison, our simulations of the Michelson response used an
instrument noise spectral density
\begin{equation}
S_n(f) = \frac{1}{4L^2}\left[4 S_{\rm pos} + 8 \left( 1 +
\cos^2(f/f_\ast)\right) \frac{S_{\rm accl}}{(2 \pi f)^4}\right]
\end{equation}
with position noise $S_{\rm pos}= 4 \times 10^{-22}\; {\rm m}^2 \,
{\rm Hz}^{-1}$ and acceleration noise $S_{\rm accl} = 9 \times
10^{-30} \; {\rm m}^2 \, {\rm s}^{-4} \, {\rm Hz}^{-1}$. The choice of
instrument noise level only has a weak effect on our results as the
unresolved galactic background is the main source of noise from
0.1~mHz to roughly 3~mHz.

A true confusion limited background is what remains after a full data
analysis procedure has removed all identifiable signals.  At present
such an algorithm has not been fully implemented, though good candidates
now exist \citep{CC05,CHR06}.  Our method of removal, by which we
remove a source using the same response approximation that included
it, mimics a true data analysis procedure, but it fails to include some
of the subtle nuances associated with signal and noise confusion
\citep{CHR06}.  Though we did not use a true source removal algorithm,
we expect that figure~\ref{fig:background_rem} represents a good
approximation to how the confusion limited background will appear in
the LISA output. The best estimates of the confusion noise may ultimately
come from Bayesian methods which treat the effective noise level as
another model parameter to be estimated \citep{UM05}.

\begin{figure}[!t]
\begin{center}
\vspace{2cm}
\includegraphics[width=0.80\textwidth]{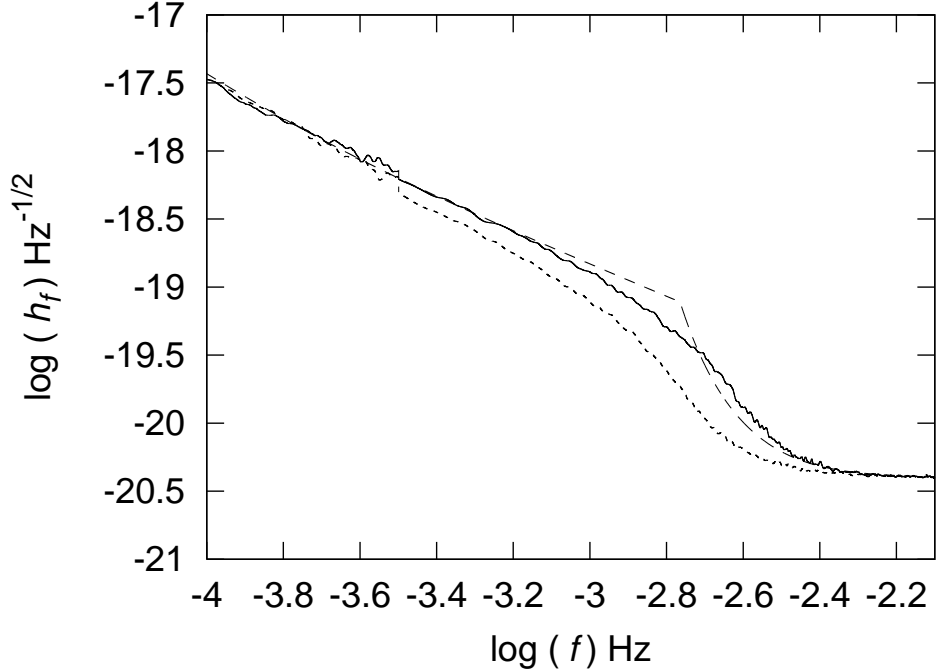}
\end{center}
\caption{Estimates of the effective noise level for the HBW 10\% model
(dotted line), and the NYZ model (solid line). The effective noise
curve used by Barack-Cutler Curve (dashed line) is also shown.}
\label{fig:confusion_noise_levels}
\end{figure}

Figure~\ref{fig:confusion_noise_levels} shows the effective noise levels
(confusion + instrument noise) for the HBW 10\% and the NYZ models. These
estimates were produced using the conservative ${\rm SNR}=10$ criteria for
bright source subtraction, and a maximum density of 0.25 resolved sources
per bin. In contrast to what we found with the HBW 100\% and HBW 10\%
(${\rm SNR}\geq 5$) models, the effective noise levels are little changed
by the source density cut-off. Also shown is the effective noise estimate
used by Barack and Cutler~\citep{BC04}. The Barack-Cutler curve agrees quite
well with our NYZ curve, though it does slightly overestimate the noise level
between 0.5 and 2 mHz.

\section{BRIGHT SOURCE STATISTICS} \label{sec:bright_sources}

The bright sources represent signals that are identifiable in LISA's
output.  By understanding their location and separation (in the
frequency domain) proper data analysis tools can be developed and
applied in the search for their signals in the detector output.
Also of interest are the properties of the bright sources.  Are they
near to us or distributed throughout the galaxy?  Can we identify the
type of binary system?  The next two sections address these issues.

\subsection{Bright Source Density}

\begin{figure}[!t]
\begin{center}
\includegraphics[width=0.80\textwidth]{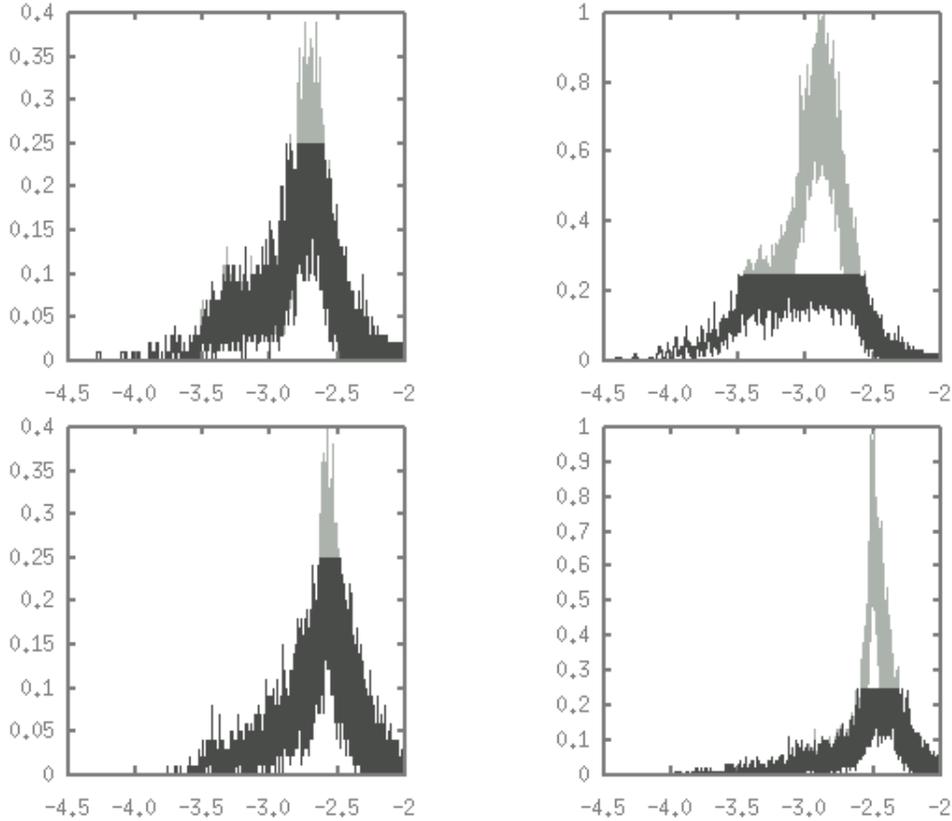}
\end{center}
\caption{The accumulated number of bright sources per frequency bin
  using a 100 bin average, both with (black) and without (grey) a
bright source density cut-off.  Starting in the upper left and going
  clockwise we have: HBW 10\% with ${\rm SNR} \geq 10$,
HBW 10\% with ${\rm SNR} \geq 5$, HBW 100\% with ${\rm SNR} \geq 10$,
and NYZ with ${\rm SNR} \geq 10$.}
\label{fig:outlier_density}
\end{figure}

For issues concerning data analysis, an interesting quantity to know
is the number of bright sources per frequency bin.
Figure~\ref{fig:outlier_density} is a plot of the average number of
bright sources per frequency bin using a 100 bin window to calculate
the average.
The peak densities occur at $\sim$2~mHz for the HBW 10\% and NYZ
models, whereas the HBW 100\% model has a maximum at $\sim$3~mHz.
For the HBW 100\% (${\rm SNR} \geq 10$) model and the 
HBW 10\% (${\rm SNR} \geq 5$) model the maximum bright source
density reached one source per bin. This is why the resolved source
density cut-off of 0.25 made such a big difference in those cases.
The bandwidth of a typical source at this frequency is approximately
twenty frequency bins (for one year of observation).  As a result, in
the peak density region there are bright sources whose power at least
partially overlaps.

\subsection{Bright Source Characteristics}

An interesting question to ask is what property makes a particular
binary bright?  At high frequencies, where the number of sources per
bin is less than unity, a source is considered bright if its signal is
greater than the intrinsic detector noise.  However, at low
frequencies, where the number of sources per bin can be in the
thousands (see fig.~\ref{fig:num_per_bin}), each source must compete
against the other sources in the bin to become detectable.

To see what makes a source bright at low frequencies recall the
functional form of the intrinsic amplitude,
\begin{equation} \label{eq:amplitude}
  \mathcal{A} = \frac{2 G^{2} M_{1} M_{2}}{c^{4} r} \left(\frac{4 \pi^2 
    f_{orb}^2}{G(M_{1}+M_{2})} \right)^{1/3} \,.
\end{equation}
Given a particular frequency bin the orbital frequency does not vary
by more than a bin width, leaving only the masses and the distance to
the source to determine if a signal is bright.

\begin{figure}[!t]
\begin{center}
\includegraphics[width=0.80\textwidth]{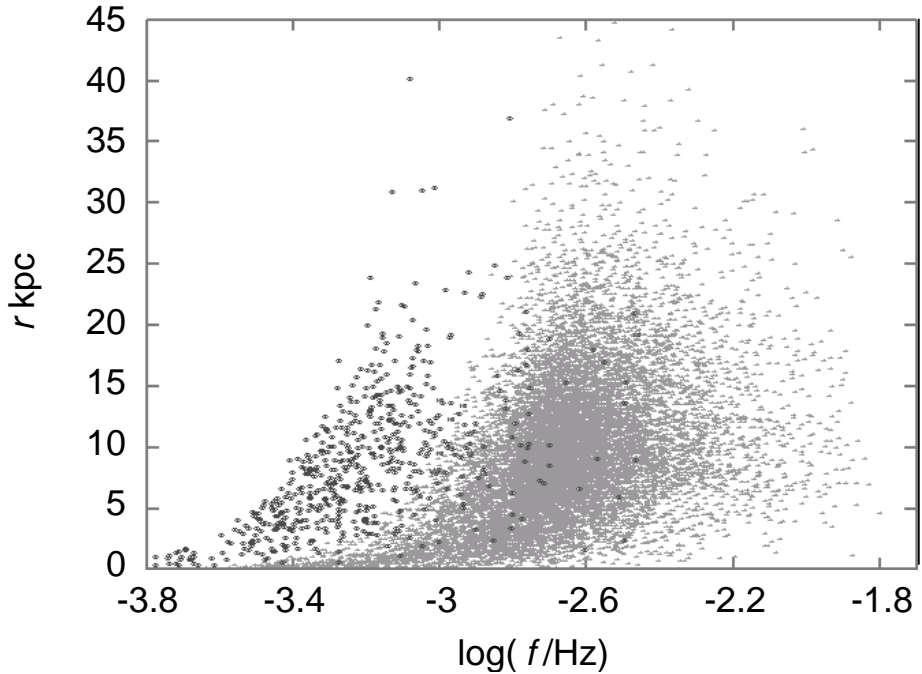}
\end{center}
\caption{Distribution of outlier distances as a function of their
  frequency.  Filled squares are white dwarf binaries, open circles
  are black hole - neutron star binaries, asterisks are are neutron
  star binaries, and open squares are cataclysmic binaries.}
\label{fig:outlier_distances}
\end{figure}

Figure~\ref{fig:outlier_distances} shows the distance versus frequency
for each bright binary coded by the type of system it is.
Evident from this figure is the mass segregation associated with the
bright systems.  At low frequencies, where the number of sources per
bin peaks, the more massive black hole - neutron stars are detectable.
At higher frequencies the black hole - neutron star binaries are less
numerous (see fig.~\ref{fig:sources_barycenter}) and the white dwarf
binaries dominate the list of bright sources.

To associate a binary to a particular type requires estimating the
component masses.  Unfortunately for sources whose frequency evolution
is too small to detect, there is a mass-distance degeneracy that
prevents direct mass measurements \citep{Cutler98}.
%Table~\ref{tab:removed_types}
Table 3 shows the number of the bright sources
that evolve by a measurable amount.  The frequency evolution is
considered measurable if the change in gravitational wave frequency during
the time of observation $T_{\rm obs}$ is greater than the width of a
frequency bin ($1/T_{\rm obs}$).

\begin{deluxetable}{cccccccc}
\tablewidth{0pt}
\tablecolumns{8}
\tablecaption{Number of Subtracted Binaries by Type}
\label{tab:removed_types}
\tablehead{
\colhead{} & 
\colhead{} & 
\multicolumn{2}{c}{Realization 1} &
\multicolumn{2}{c}{Realization 2} &
\multicolumn{2}{c}{Realization 3} \\
\cline{3-4}
\cline{5-6}
\cline{7-8}
\colhead{Type} &
\colhead{Included} &
\colhead{Removed} &
\colhead{Evolving} &
\colhead{Removed} &
\colhead{Evolving} &
\colhead{Removed} &
\colhead{Evolving}}
\startdata
W~UMa & $3 \times 10^{7}$ & 0 & 0 & 0 & 0 & 0 & 0 \\
CB & $1.8 \times 10^{6}$ & 1 & 0 & 2 & 0 & 1 & 0 \\
NS - NS & $10^{6}$ & 11 & 0 & 11 & 2 & 7 & 0 \\
BH - NS & $5.5 \times 10^{5}$  & 591 & 10 & 585 & 12 & 550 & 8 \\
WD - WD & $3 \times 10^{6}$  & 13,302 & 346 & 13,681 & 353 & 13,539 & 337 \\
\hline
Total & $3.635 \times 10^{7}$ & 13,905 & 356 & 14,279 & 367 & 14,097 & 345 \\
\enddata
\end{deluxetable}

It is interesting to note that very few sources will have a measurable
frequency evolution during one year of observation.  The two main
reasons for this is that the frequency evolution is dictated by mass
and initial frequency (see eq.~[\ref{eq:period_evolution}]).  The
massive black hole - neutron star binaries are located at low
frequencies, while the higher frequency white dwarf binaries have
smaller chirp masses.

Figure~\ref{fig:outlier_distances} also demonstrates that we should be
able to see individual binaries distributed throughout the galaxy.
This introduces the tantalizing prospect of using gravitational wave
data to map galactic populations.  However, since there is a
mass-distance degeneracy for systems that do not evolve appreciable
during the lifetime of the detector, only a small fraction of the
identifiable sources can be used for such an analysis.

\section{SUMMARY AND CONCLUSIONS} \label{sec:conclusion}

In this paper we have presented a Monte Carlo simulation of the
galactic gravitational wave background as it would be detected by the
proposed LISA mission after one year of operation.  For the
intrinsic binary properties we used
the distributions given in \citet{HBW90} and \citet{NYZ04}.  Our key
findings are: the
galactic background levels will be reduced in the detector frame as
compared to the barycenter frame, prior to the removal of the bright
sources the background is not characterized by a Gaussian
distribution, and of the $\sim$$10^{4}$ identifiable sources only
$\sim$$10^{2}$ are evolving and thus identifiable by type. We have
also derived a new estimate for the confusion limited background that
differs from other estimates given in the
literature. Below 1~mHz our estimate is lower than previous estimates,
while above 1~mHz our estimate is higher than previous findings. Our
calculation of the bright source density (in frequency space) suggests
that data analysis algorithms can be developed that are capable of
resolving $\sim$10,000 galactic binaries from a one year LISA data
stream. Of these, roughly three hundred will be measurably chirping,
allowing the determination of the chirp mass and the distance to these
sources.  The number of resolvable galactic sources, especially the
number of measurably evolving systems, will increase significantly
after several years of observation (the resolution of the frequency
derivative improves with observation time, $T_{\rm obs}$,
as $T^{5/2}_{\rm obs}$).

An important point to keep in mind about the results presented here is
that they assumed particular descriptions for the galactic
distribution, total number, and source characteristics of each binary
population.  A different collection of models of the extrinsic
parameters may return a slightly different set of results as is seen
in the NYZ backgrounds.  However, the main conclusions drawn here
are largely dictated by three quantities, the total number of binary
systems, their period distributions, and the component masses.

The number of binary sources in the galaxy can impart a noticeable
difference in the background levels.  As radiation from the binaries
converges on the detector, the random phase differences will cause
constructive and destructive interference.  Statistically the problem
is analogous to a random walk.  As a result, the net spectral
amplitude per frequency bin will grow as $\sqrt{N}$.  For every factor
of one hundred difference in the number of sources, the background
levels raise or lower by a decade respectively.  For
electromagnetically visible binaries (W~UMa, unevolved, and
cataclysmic binaries) galactic surveys have placed stringent
constraints on the total number of such binaries.  However, for the
compact binaries (neutron star - neutron star, black hole - neutron
star, and white dwarf binaries) equivalent surveys have failed to
place strict bounds on the total numbers.  One return of the LISA
mission will be to place limits on the populations by measuring the
galactic background median levels and the number of bright sources
from each population.

While the overall level for the galactic background level is partially
influenced by the total number of systems, the background level is
also dictated by the component masses via
equation~\eqref{eq:amplitude}.  By comparing our HBW based simulations
to the population synthesis approach of \citet{NYZ04}, we have found that
relatively small changes in the component masses can lead to significant
changes in the background levels. Current uncertainties in
the true mass distributions arise form a lack a observational data and
a theoretical understanding of mass transferring stages during
formation. The NYZ white dwarf binaries
are typically composed of two light components, $M < 0.5 M_{\odot}$, whereas
the HBW white dwarf binaries are typically composed of one light
component and one heavy component, $M > M_{\odot}$. This results in
the HBW systems having mean chirp masses,
${\cal M}=(M_1 M_2)^{3/5}/(M_1+M_2)^{1/5}$, almost a factor of two higher
than the NYZ systems. Since the amplitude of signals scales as
${\cal M}^{5/3}$, the difference in component masses translates into
a factor of $\sim 3.2$ increase in the HBW background relative to the
NYZ background for the same source density. It is an interesting
numerical coincidence that the 10\% reduction in source density
proposed by HBW yields a $\sqrt{10} \sim 3.2$ reduction in the
amplitude of the background, resulting in a background level comparable
to the NYZ model. Despite this fortuitous cancellation, the HBW 10\%
and the NYZ models yield different confusion noise estimates due
to the reduced number of sources in the HBW 10\% model.

The period distributions will also impart a noticeable change in the
background.  If, for example, there is a mechanism that suppresses low
period white dwarfs, the values in table~\ref{tab:removed_types} and
figure~\ref{fig:outlier_distances} would change.  Conversely, we can
invert the problem and ask questions such as what would the high
frequency end of the galactic background look like if certain physics
are included in the models for close white dwarf binary production?

%==== Acknowledgments =============================

\acknowledgments

We are very grateful to Gils Nelemans for providing us with a realization
of his white dwarf background.
We would like to thank Peter Bender for helpful discussions
concerning the confusion limited background.  This work was supported
by the NASA EPSCoR program through Cooperative Agreement NCC5-579.
LJR and SET acknowledge the support of the Center for Gravitational
Wave Physics. The Center for Gravitational Wave Physics is supported
by the NSF under Cooperative Agreement PHY 01-14375. SET is also supported
by a Pennsylvania State University Graduate Fellowship.

%==== Appendix ====================================

\appendix

\section{DETECTOR RESPONSE}

For a gravitational wave traveling in the $\hat{k}$ direction we can
express a single channel LISA response as a linear sum of responses
for each polarization state,
\begin{equation} \label{eq:response}
s(t) = A_{+} F^{+}(t) \cos\Phi(t) + A_{\times} F^{\times}(t) \sin
\Phi(t)\,,
\end{equation}
where the wave phase is given by
\begin{equation}
\Phi(t) =  2\pi f_{o} + \pi \dot{f}_{o} t^2 + \varphi_{o} -
\Phi_D(t) \,. 
\end{equation}
Here $f_{o}$ is the initial gravitational wave frequency,
$\dot{f}_{o}$ is the initial frequency derivative, and $A_{+,\times}$
are the polarization amplitudes given in
equation~\eqref{eq:pol_amps}. The Doppler modulation of the signal is
given by
\begin{equation}
\Phi_D(t) \simeq \frac{2\pi f_{o}}{c} \, \hat{k} \cdot \bs{x}_{i}(t)
\,.
\end{equation}
where we have neglected the small correction due to the frequency
evolution $\dot{f}_{o}$.  The antenna beam pattern functions
$F^{+,\times}(t)$ describe LISA's time varying sensitivity to each
polarization and are given by
\begin{mathletters} \label{eq:FpFc}
\begin{eqnarray}
F^{+}(t) &=& \frac{1}{2} \Big( \cos(2\psi) D^{+}(t) - \sin(2\psi)
D^{\times}(t) \Big) \\
F^{\times}(t) &=& \frac{1}{2} \Big( \sin(2\psi) D^{+}(t)+ \cos(2\psi)
D^{\times}(t) \Big) \,,
\end{eqnarray}
\end{mathletters}
where the two-arm detector response functions are defined as
\begin{mathletters}\label{eq:DpDc}
\begin{eqnarray}
D^{+}(t) &\equiv& d^+_{ij}\mathcal{T}_{ij}(t,f_{gw})
-  d^+_{ik}\mathcal{T}_{ik}(t,f_{gw})  \\
D^{\times}(t) &\equiv& d^\times_{ij}\mathcal{T}_{ij}(t,f_{gw})
-  d^\times_{ik}\mathcal{T}_{ik}(t,f_{gw})  \\
\end{eqnarray}
\end{mathletters}
with
\begin{mathletters}
\begin{eqnarray}
d^{+}_{ij}(t) &\equiv& \left( \hat{r}_{ij}(t) \otimes \hat{r}_{ij}(t)
\right) : \bsrm{e}^{+} \\
d^{\times}_{ij}(t) &\equiv& \left( \hat{r}_{ij}(t) \otimes
\hat{r}_{ij}(t) \right)  :\bsrm{e}^{\times} \,.
\end{eqnarray}
\end{mathletters}
The colon denotes a double contraction, $\bsrm{x}:\bsrm{y} = x^{ab}
y_{ab}$, with repeated indices implying a summation, $\hat{r}_{ij}(t)$
is a unit vector that points from spacecraft $i$ to spacecraft $j$,
and $\mathcal{T}_{ij}(t,f_{gw})$ is the round-trip transfer function
for the arm connecting the $i$ and $j$ spacecraft,
\begin{eqnarray} \label{eq:transfer}
\mathcal{T}_{ij}(t,f) &=& \frac{1}{2} \left[ \sinc \left(
\frac{f}{2f_{\ast}} \left(1 - \hat{k} \cdot \hat{r}_{ij}(t) \right)
\right) \exp \left( -i \frac{f}{2f_{\ast}} \left(3 + \hat{k} \cdot
\hat{r}_{ij}(t) \right) \right) \right. \nonumber\\
&& + \left. \sinc \left(\frac{f}{2f_{\ast}}\left(1 + \hat{k} \cdot
\hat{r}_{ij}(t) \right) \right) \exp \left(-i \frac{f}{2f_{\ast}} \left( 1 +
\hat{k} \cdot \hat{r}_{ij}(t) \right) \right) \right]
\end{eqnarray}
The quantity $f_{\ast} \equiv c/2\pi L$ is referred to as the transfer
frequency, which for LISA ($L = 5 \times 10^{6}$~km) has a value of
9.54~mHz.  The transfer frequency is approximately the point at which
a gravitational wave will ``fit inside'' the detector arms.

The polarization basis tensors are expressed in terms of two
orthonormal unit vectors,
\begin{mathletters}
\begin{eqnarray}
\bsrm{e}^{+} &=& \hat{u} \otimes \hat{u} - \hat{v} \otimes \hat{v} \\
\bsrm{e}^{\times} &=& \hat{u}\otimes\hat{v}+\hat{v}\otimes\hat{u} \,.
\end{eqnarray}
\end{mathletters}
The unit vectors $\hat{u}$ and $\hat{v}$, along with the propagation
direction of the gravitational wave, $\hat{k}$, form an orthonormal
triad, which may be expressed as functions of the source location on
the celestial sphere ($\theta, \phi$),
\begin{mathletters}
\begin{eqnarray}
\hat{u} &=& \cos(\theta)\cos(\phi)\hat{x} +
\cos(\theta)\sin(\phi)\hat{y} - \sin(\theta)\hat{z} \\
\hat{v} &=& \sin(\phi)\hat{x} - \cos(\phi)\hat{y} \\
\hat{k} &=& -\sin(\theta)\cos(\phi)\hat{x} -
\sin(\theta)\sin(\phi)\hat{y} - \cos(\theta)\hat{z} \,.
\end{eqnarray}
\end{mathletters}

To calculate the unit vectors $\hat{r}_{ij}(t)$ we use the spacecraft
coordinates given by \citet{CR03},
\begin{mathletters} \label{eq:positions}
\begin{eqnarray}
x(t) &=& R \cos(\alpha) + \frac{1}{2} \varepsilon R \big( \cos(2\alpha
- \beta) - 3\cos(\beta) \big) \\
y(t) &=& R \sin(\alpha) + \frac{1}{2} \varepsilon R \big( \sin(2\alpha
- \beta) - 3\sin(\beta) \big) \\
z(t) &=& -\sqrt{3} \varepsilon R \cos(\alpha - \beta) \,.
\end{eqnarray}
\end{mathletters}
In the above $R = 1$~AU is the orbital radius of the guiding center,
$\varepsilon$ is the orbital eccentricity, $\alpha \equiv 2 \pi
f_{m} t + \kappa$ is the orbital phase of the guiding center, $f_{m} =
1/\textrm{year}$ is the modulation frequency, and $\beta \equiv 2\pi
(n-1)/3 + \lambda$ ($n = 1,2,3$) is the relative phase of the
spacecraft within the constellation.  The parameters $\kappa$ and
$\lambda$ give the initial ecliptic longitude and orientation of the
constellation respectively.  Note that to linear order in the
eccentricity, which is the order we work to, the triangular formation
is rigid with arm lengths given by $L = 2 \sqrt{3} \varepsilon R$.

Equation~\eqref{eq:response}, and the relationships that follow it,
represent the \textit{Rigid Adiabatic Approximation} described in
\citet{RCP04}. An important property of the approximation is that it
is implemented in the time domain.  To properly model sources with
frequencies up to 10~mHz requires a minimum of $\sim\!\!10^{6}$ data
points for an observational period of one year. The computational cost of
simulating the response for over $10^7$ sources in the time domain is
prohibitive.

A desirable alternative is to work directly in the frequency domain.
The advantage of doing so is that the number of relevant Fourier
coefficients is small.  Here the concept of ``relevant'' are those
coefficients that contain a high percentage ($\sim\!\!98\%$) of the
spectral power.  For reference, a moderate signal-to-noise ratio
source at 10~mHz, observed for one year, will spread across sixty-five
frequency bins. It is possible to derive an analytic expression for
the Fourier transform of the time domain signal,
equation~\eqref{eq:response}.  The calculation has four steps. First,
the amplitude and frequency modulations are decomposed into harmonics
of the detector's orbital frequency $f_m$. Second, the frequency
evolution term, $\exp(\pi i \dot{f}_{o} t^2)$, is Fourier transformed.
Third, the product of the orbital harmonics and the barycenter wave
function $\exp(2\pi i f_{o} t)$ are Fourier transformed, yielding a
Fourier series who's coefficients are products of the harmonic
amplitudes and the cardinal sine function. Finally, the Fourier series
from steps two and three are convolved to give the complete finite
time Fourier transform of the time domain signal.  Our calculation
generalizes the expression derived by \citet{CL03b} to allow for
arbitrary observation times, chirping sources, and the effect of the
instrument transfer functions.

The first step in the calculation relies on the fact that the
functions $F^+(t)$, $F^\times(t)$, and $\Phi_D(t)$ owe their time
variation to the orbital motion of the detector. Thus we may decompose
each of these functions into harmonics of the orbital frequency $f_m$,
\begin{mathletters}
\begin{eqnarray}
F^{+}(t) &=& \sum_{n} p_{n} e^{2\pi i n f_m t} \\
F^{\times}(t) &=& \sum_{n} c_{n} e^{2\pi i n f_m t} \\
e^{i\Phi_D(t)} & = & \sum_{n} d_{n} e^{2\pi i n f_m t} \,.
\end{eqnarray}
\end{mathletters}
The coefficients $d_n$ are given by the Jacobi-Anger expansion,
\begin{equation}
d_{n} = J_{n}(2 \pi f (R/c) \sin\theta)e^{in(\pi/2-\phi)} \, ,
\end{equation}
where $J_n$ is the Bessel function of the first kind of order $n$.
Deriving expressions for $p_n$ and $c_n$ is complicated by the transfer
functions that appear in equation \eqref{eq:DpDc}. While it is
possible to perform the harmonic decomposition exactly using the
Jacobi-Anger expansion, a simpler approach is to Taylor expand the
transfer functions in powers of $f/f_{\ast}$ then decompose each term
into orbital harmonics. For our current needs a second order expansion
is sufficient,
\begin{eqnarray}
\mathcal{T}_{ij}(t,f) &=& 1 - i \left(2 + \hat{k}\cdot\hat{r}_{ij}(t)
\right) \left( \frac{f}{2 f_{\ast}} \right) \nonumber\\
&& - \frac{1}{2} \left(4 + 3\left( 2 + \hat{k}\cdot\hat{r}_{ij}(t)
\right) + \left( 2 + \hat{k}\cdot\hat{r}_{ij}(t)\right)^2 \right)
\left( \frac{f}{2 f_{\ast}} \right)^{2} + \mathcal{O}\left(\frac{f}{2
f_{\ast}} \right)^{3} \,.
\end{eqnarray}
We can now re-express $\mathcal{T}_{ij}$, $d^{+}_{ij}$, and
$d^{\times}_{ij}$ in terms of orbital harmonics:
\begin{eqnarray}
\mathcal{T}_{ij}(t,f) & = & \sum_{n=-4}^{4} \widetilde{\mathcal{T}}_{ij,n}
e^{2 \pi i n f_m t} \nonumber\\
d^{+}_{ij}(t,f) & = & \sum_{n=-4}^{4} \widetilde{d^{+}}_{ij,n}
e^{2 \pi i n f_m t} \nonumber\\
d^{\times}_{ij}(t,f) & = & \sum_{n=-4}^{4} \widetilde{d^{\times}}_{ij,n}
e^{2 \pi i n f_m t} \, .
\end{eqnarray}
Convolving these expansions yields
\begin{mathletters}
\begin{eqnarray}
p_{n} &=& \sum_{l=-4}^{4} \sum_{m=-4}^{4} \left(
\widetilde{d}_{ij,l}^{+} \widetilde{\mathcal{T}}_{ij,m} -
\widetilde{d}_{ik,l}^{+} \widetilde{\mathcal{T}}_{ik,m} \right) \\
c_{n} &=& \sum_{l=-4}^{4} \sum_{m=-4}^{4} \left(
\widetilde{d}_{ij,l}^{\times} \widetilde{\mathcal{T}}_{ij,m} -
\widetilde{d}_{ik,l}^{\times} \widetilde{\mathcal{T}}_{ik,m} \right)
\,,
\end{eqnarray}
\end{mathletters}
where $n = l+m$.  The range of the sums in the above harmonic
decomposition can be traced to the functional form of the
$\hat{r}_{ij}(t)$ unit vectors.  From equation~\eqref{eq:positions} we
see that the harmonic decomposition of $\hat{r}_{ij}(t)$ involves a
sum from -2 to 2.  The one-arm detector response functions and the
expanded transfer functions are each a function of $\hat{r}^{2}$ and,
therefore, their decompositions range from -4 to 4.

The second step is to Fourier transform the frequency evolution term,
which yields Fourier coefficients $q_{n}$ that can be expressed in
terms of error functions of a complex argument. The third step is to
Fourier transform the remaining terms over the finite observation time
$T$ which has the effect of introducing the cardinal sine function,
$\sinc(x) \equiv \sin(x)/x$. Putting everything together we have
\begin{equation}\label{Full_Signal_Chirp}
s_{j} = \frac{1}{2} e^{i\varphi_{0}}\sum_{k} q_{k} \sum_l
\sinc(x_{lm}) e^{i x_{lm}} \sum_{n} \left(A_{+} p_{n} + e^{i 3\pi/2}
A_{\times} c_{n} \right) \sum_{p} d_{p}\,,
\end{equation}
where $x_{lm} = \pi(l f_m  + f_0 T - m)$, $j = k +l$ and $m=n+p$.

\begin{figure}[!t]
\begin{center}
\includegraphics[width=0.80\textwidth]{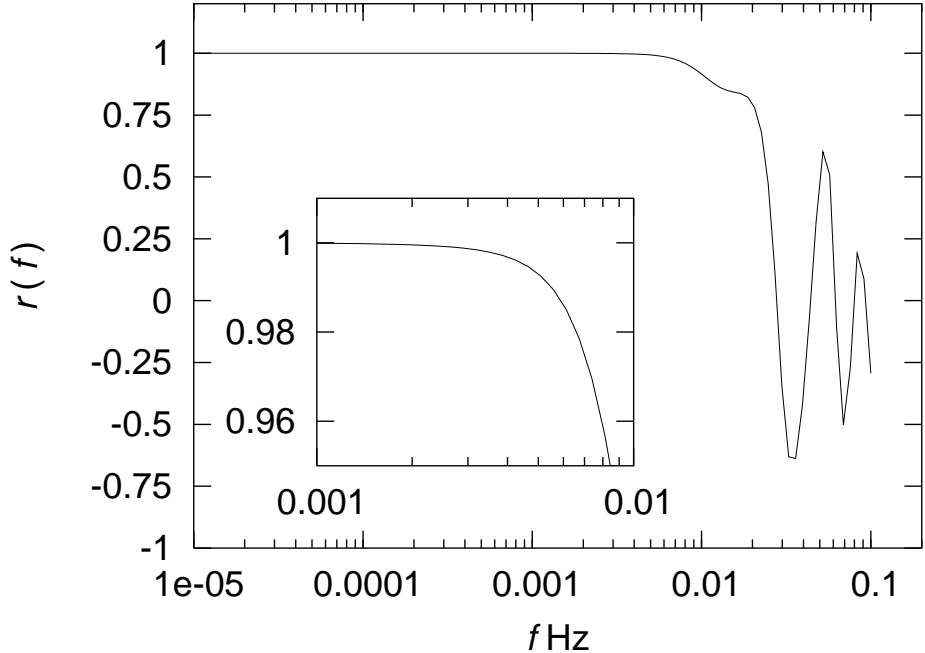}
\end{center}
\caption{The correlation between the \textit{Extended Low Frequency
    Approximation} and the \textit{Rigid Adiabatic Approximation}.}
\label{fig:correlation}
\end{figure}

To test the range of validity for the new approximation, referred to
as the \textit{Extended Low Frequency Approximation} (ELF), we
calculated a normalized correlation between the new approximation and
the \textit{Rigid Adiabatic Approximation} (RA),
\begin{equation}
r(f) = \frac{\langle s_{RA}(f) | s_{ELF}(f) \rangle}{ \sqrt{\langle
s_{RA}^2(f) \rangle \langle s_{ELF}^2(f)} \rangle} \,.
\end{equation}
The rigid adiabatic adequately describes LISA's response up to
$\sim$0.3~Hz \citep{RCP04} making it a valid benchmark to compare the
new approximation against.  Figure~\ref{fig:correlation} shows the
correlation between the two approximations for a randomly selected
binary who's masses, sky location, etc., were held fixed while the
correlation was tested at different frequencies.
To safely apply the \textit{Extended Low Frequency Approximation}, in
our analysis we use a cutoff of 7~mHz as the trigger point for where
we switched to the slower \textit{Rigid Adiabatic Approximation}.
This allows for quick and accurate frequency domain modeling for all
but $\sim$100 sources in our galactic simulations.

%==== Bibliography ================================

\end{document}